%% file: plasmaring_new.tex
\documentclass[a4paper,10pt]{article}

\usepackage{a4wide}
\usepackage[centertags]{amsmath}
\usepackage{amssymb}
\usepackage[sort&compress,numbers]{natbib}
\usepackage{ifpdf}
\usepackage{pgf}
\ifpdf
\usepackage[pdftex,bookmarks]{hyperref}
\else
\usepackage[hypertex]{hyperref}
\fi
\input{mydefs.tex}

\input{newsymb.tex}
\newcommand{\tloc}{\mathcal{T}}

\newcommand{\ploc}{\mathcal{P}}

\newcommand{\gpf}{\mathcal{Z}_\mathrm{gc}}
\newcommand{\mfp}{l_{\mathrm{mfp}}}
\newcommand{\eos}{\alpha}
\newcommand{\tc}{\mathcal{T_\mathrm{c}}}
\newcommand{\rz}{\rho_0}
\newcommand{\rc}{\rho_\mathrm{c}}

\newcommand{\Ra}{R_\text{avg}}
\newcommand{\tE}{\widetilde{E}}
\newcommand{\tL}{\widetilde{L}}
\newcommand{\tS}{\widetilde{S}}
\newcommand{\tT}{\widetilde{T}}
\newcommand{\tw}{\widetilde{\omega}}
\newcommand{\tgpf}{\widetilde{\mathcal{Z}}_\mathrm{gc}}

\title{Lumps of plasma in arbitrary dimensions}
\author{Jyotirmoy Bhattacharya$^{(a)}$ and Subhaneil Lahiri$^{(b)}$\\
\small{\emph{$^{(a)}$Department of Theoretical Physics,
                   Tata Institute of Fundamental Research,}}\\
\small{\emph{Homi Bhabha Rd, Mumbai 400005, India}}\\
\small{\emph{$^{(b)}$Jefferson Physical Laboratory,
                   Harvard University, Cambridge MA 02138, USA}}
}

\begin{document}

\maketitle
\begin{abstract}
We use the AdS/CFT correspondence in a regime in which the field theory reduces to
fluid dynamics to construct an infinite class of new black objects in Scherk-Schwarz compactified
AdS$_{d+2}$ space. Our configurations are dual to black objects that generalize black rings and have
horizon topology  $S^{d-n} \times T^n$ for $n \leq \frac{d-1}{2}$. Locally our fluid configurations
are plasma sheets that curve around into tori whose radii are large compared to the thickness of the sheets
(the ratio of these radii constitutes a small parameter that permits the perturbative construction
of these configurations). These toroidal
configurations are stabilized by angular momentum. We study solutions whose dual horizon
topologies are $S^3\times S^1$, $S^4\times S^1$ and $S^3\times T^2$ in detail; in particular we
investigate the thermodynamic properties of these objects.  We also present a formal general construction of
the most general stationary configuration of fluids with boundaries that solve the $d$ dimensional relativistic Navier-Stokes equation.
\end{abstract}

\tableofcontents

\section{Introduction}

In this paper we attempt to understand horizon topologies and thermodynamics
of black objects in arbitrary high dimensional Scerk-Schwarz compactified
AdS spaces (SSAdS). The spectrum of black objects in more than 4 dimensions
is extremely rich and consequently has drawn considerable interest recently
\cite{Emparan:2007wm,Niarchos:2008jc}. As the construction of these exotic horizon topologies
directly in gravity turns out to be technically difficult we study them in a
somewhat indirect manner using the AdS/CFT correspondence
\cite{Maldacena:1997re,Gubser:1998bc,Witten:1998qj}.

This correspondence relates a theory of gravity in AdS$_{d+2}$ to a CFT in
$d+1$ dimensional spacetime. We consider the field theory obtained by
Scherk-Schwarz compactification of this dual CFT, which consequently lives in
$d$ dimensions. This field theory has a first order confinement/deconfinement
phase transition. This corresponds to a Hawking-Page-like phase transition in
the bulk, for which the low temperature phase is the AdS-soliton and the high
temperature phase is a large AdS black brane \cite{Witten:1998zw}.

In the long wavelength limit, this field theory admits a fluid
description where the dynamics is governed by the $d$ dimensional
relativistic Navier-Stokes equation. The effect of the Scerk-Schwarz
compactification is only to introduce a constant additive piece to the free
energy of the deconfined fluid \cite{Aharony:2005bm}. Due to this shift, the
pressure can go to zero at finite energy densities, allowing the existence of
arbitrarily large finite lumps of deconfined fluid separated from the
confined phase by a surface -- the plasmaballs of  \cite{Aharony:2005bm}.
Now by the AdS/CFT correspondence finite energy
localized non-dissipative configurations of the plasma fluid in the
deconfined phase is dual to stationary black objects in the bulk. Thus, by
studying fluid configurations that solve the $d$ dimensional relativistic
Navier-Stokes equation we can infer facts about the black objects in
SSAdS$_{d+2}$ \cite{Lahiri:2007ae,Bhardwaj:2008if}.

Two important feature of the dual black object that one can infer from the
fluid configurations are the horizon topology and the thermodynamics. The
thermodynamics of the black object can be studied by simply computing the
thermodynamic properties of the fluid configuration -- one integrates the
energy density, entropy density etc.\ to compute the total energy, entropy
etc.\ and the rest follows.

\begin{figure}[htbp]
  \begin{center}
  \input{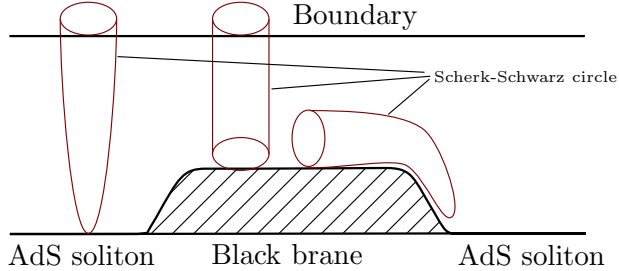}
  \end{center}
  \caption{Schematic description of the bulk dual of a plasmaball with some circle fibres indicated.}\label{fig:circlefibre}
\end{figure}

The horizon topology can be inferred as follows. Far outside the region
corresponding to the plasma, the bulk should look like the AdS-soliton. In
this configuration the Scherk-Schwarz circle contracts as one moves away from
the boundary, eventually reaching zero size and capping off spacetime
smoothly. Deep inside the region corresponding to the plasma, the bulk should
look like the black brane. In this configuration the Scherk-Schwarz circles
does not contract, it still has non-zero size when one reaches the horizon.
It follows that as one moves along the horizon, the Scherk-Schwarz circle
must contract as one approaches the edge of the region corresponding to the
plasma. The horizon topology is found by looking at the fibration of a circle
over a region the same shape as the plasma configuration, contracting the
circle at the edges \cite{Aharony:2005bm, Lahiri:2007ae}. We have provided a
schematic drawing of this in fig.\ref{fig:circlefibre}.

In the fluid description the degrees of freedom includes the velocity field,
$u^{\mu}(x)$, and the temperature field, $\tloc(x)$, (we consider uncharged
fluids dual to uncharged black objects; otherwise the degrees of freedom
would also include the chemical potentials for those charges). Now as we
seek time-independent solutions, Lorentz symmetry allows us to consider
fluid velocities of the form
\begin{equation*}
 u^{\mu} = \gamma (\partial_t + \omega_a l_a) ,
\end{equation*}
where $l_a$ are the killing vectors along the Cartan direction of the spatial
rotation group, $\gamma$ is the normalization and the $\omega_a$ are some
constants. This along with the fact that our solutions are non-dissipative
forces the temperature field to be of the form
\begin{equation*}
 \tloc = \gamma T,
\end{equation*}
where $T$ is a constant. With a simple thermodynamic argument we show that
$T$ is the overall thermodynamic temperature of the fluid configuration and
$\omega_a$ are the thermodynamic angular velocities. Further we demonstrate
that the equations of motion for non-dissipative time-independent solutions
at the surface of the fluid configuration reduce to the condition
\begin{equation*}
 \ploc|_\text{surface} = \sigma \Theta,
\end{equation*}
where $\sigma$ is the surface tension and $\Theta$ is the trace of the
extrinsic curvature of the fluid surface under consideration. The pressure
$\ploc$ is related to the temperature $\tloc$ by the equation of state, so
this provides a differential equation for the position of the surface. These
configurations are parameterized by the temperature $T$ and the angular
velocities $\omega_a$.

We then proceed to construct a class of fluid configurations whose surface is
a solution of the above equation in a certain limit. In $d$ spacetime
dimensions the topologies of these configurations are
\begin{equation*}
 B^{(d-1-n)} \times T^n = B^{(d-1-n)} \times \underbrace{ S^1 \times S^1 \dots  \dots S^1}_{\text{n times}},
\end{equation*}
where $n$ satisfies
\begin{equation*}
 \begin{split}
  n &= 0 , \quad \text{for } d=3,\\
  n &\leq \frac{d-1}{2}, \quad \text{for odd } d \text{ greater than } 3,\\
  n &\leq \frac{d-2}{2}, \quad \text{for even d}.
 \end{split}
\end{equation*}
These solutions are rotating in the plane in which the $S^1$s lie, and for
simplicity we turn off angular momentum along any other directions.
In these configurations pressure in the radial direction of the ball is
balanced by the surface tension. While along the radial direction of the
$S^1$s the centrifugal force balances the pressure (therefore rotation is essential
in the plane of the $S^1$s).
We refer to the limit in which the (average) radius of the ball is small compared to
the (average) radius along the $S^1$s \footnote{When there is more than one
$S^1$ this radius refers to the magnitude of the vector which is obtained by
the vector sum of the radii of the various $S^1$s} as `the generalized thin
ring' limit. The ratio of these two radii serves as the small parameter in
the problem. To leading order in this parameter we find that the fluid
configurations are exactly $ B^{(d-1-n)} \times T^n$
(in contrast to merely having the same topology). The force balance
conditions then relate the intrinsic fluid parameters (the temperature
and the angular velocities) to the parameters of the fluid configuration
(the radius of the ball and the radii of the various $S^1$s).
These fluid configurations
are dual to black objects with horizon topologies $ S^{(d-n)} \times T^n$
and hence this provides an indirect proof of
existence of such exotic horizon topologies of black objects in
SSAdS$_{d+2}$. This approach is reminiscent of (and inspired by) the
black-fold approach of \cite{Emparan:2009cs}.

Further it is possible to construct a well controlled perturbation theory
about these generalized thin ring solution. This we demonstrate by explicitly
computing the leading order corrections to the thin ring solutions in
specific examples, namely the ring in 4 dimensions and the ring and the
`torus' (the one with the topology $B^2 \times T^2$) in 5
dimensions. We find that the leading order correction only appear at the
second order in the expansion parameter (the small parameter described
above). Also in these cases we explicitly compute the thermodynamic
quantities (which are again correct up to second order in the expansion
parameter) with which we construct the phase diagrams of these solutions
within appropriate validity regimes.

The rest of the paper is organized as follows. In \S\ref{sec:general}, we
review the general formalism we will use -- the thermodynamics of the type of
confining theory we consider in this paper, relativistic fluid mechanics with
surface tension and the construction and thermodynamics of equilibrium
configurations.

In \S\ref{sec:arbdim}, we discuss the construction of configurations in
arbitrary dimensions in the `thin ring' limit to first order in the expansion
parameter. We do not take the expansion any further, as the coordinate system
we use is ill suited to perturbation theory.

We then switch to adapted coordinate systems and study three examples in
detail. In \S\ref{sec:rings} we study the ring solutions in four and five
dimensions (topologies $B^2 \times S^1$ and $B^3 \times S^1$ respectively).
We construct the solutions and study their thermodynamics to second order in
the expansion parameter. In \S\ref{sec:torus}, we do the same for the `torus'
solution in five dimensions (topology $B^2 \times T^2$).

In Appendix \ref{sec:extrinsic}, we compute some formulae for the extrinsic
curvature of the surfaces. In Appendix \ref{sec:largeexp}, we present
expressions for the energy, entropy and angular momentum of the ring and
`torus' in five dimensions that were too large to include in the main text.
We summarise the notation used in this paper in Appendix \ref{sec:notation}.

\section{Hydrodynamics and overall thermodynamics of plasma-lumps} \label{sec:general}

In this section we review the general formalism we will use in this paper. In \S\ref{sec:fltherm} we discuss the thermodynamic properties of the fluids we consider here, in \S\ref{sec:stress} we review relativistic fluid mechanics, in \S\ref{sec:surface} we review the relativistic treatment of surface tension and in \S\ref{sec:rigidrot} we describe the general construction of equilibrium configurations.

\subsection{Thermodynamics}\label{sec:fltherm}

A fluid with all conserved charges and chemical potentials set to zero
satisfies
\begin{equation}\label{inttherm:eq}
  \begin{split}
    \rho+\ploc &= s\tloc ,\\
    \dr\rho &= \tloc \dr s ,\\
    \dr\ploc &= s\, \dr\tloc ,\\
  \end{split}
\end{equation}
where $\rho$, $\ploc$, $s$ and $\tloc$ are the local density, pressure,
entropy density and temperature as measured in the rest frame of the fluid.
Note that all intensive thermodynamic quantities can be written as functions
of one variable 
which we will usually choose to be the temperature. 
Once we are given the pressure as a function of temperature, 
we can use \eqref{inttherm:eq} to determine the other quantities.

The fluids that we will consider here -- those obtained by the compactification of a conformal theory with a gravity dual on a Scherk-Schwarz circle -- have an equation of state of the form
\begin{equation}\label{bbeos:eq}
  \ploc = \frac{\eos}{\tc}\prn{\tloc^{d+1} - \tc^{d+1}}.
\end{equation}
Note that this notation is slightly different from that used in \cite{Lahiri:2007ae,Bhardwaj:2008if}. The quantity $\eos$ here differs from the one used previously by a factor of $\tc$. We also have $\rho_0 = \eos\tc^d$.

When the parent conformal theory is $\CN=4$ super Yang-Mills, we have $\eos=\frac{\pi^2 N^2}{8}$.

As the confined phase has a free energy $\sim\CO(N^0)$, to leading order at large $N$ we can treat it as the vacuum. The confining/deconfining phase transition occurs when the two phases have the same free energy. In our case, this is approximately at $\tloc=\tc$. At this temperature, the density is given by $\rc=(d+1)\eos\tc^d$. Note that $\rho_0$ is not the critical density.

\subsection{Fluid mechanics}\label{sec:stress}


Provided all length scales are large compared to the thermalisation scale of the fluid (which we call $\mfp$), each patch of the fluid is well described by equilibrium thermodynamics in its rest frame. The fluid is characterised by the velocity of these patches --- described by a vector $u^\mu=\gamma(1,\vec{v})$ --- and the intensive thermodynamic quantities in their rest frames --- which can all be computed from the proper temperature $\tloc$ 
using the equation of state and the first law of thermodynamics, as in \S\ref{sec:fltherm}.

The equations of fluid dynamics are simply a statement of the conservation of the stress tensor $T^{\mu \nu}$ 
\begin{equation}\label{Epconsv:eq} \begin{split}
  \nabla_\mu T^{\mu\nu} &
                        = 0 .
\end{split}
\end{equation}
%
This provides $d$ equations for the evolution of for the $d$ quantities $\vec{v}$ and $\tloc$ once we have expressed the stress tensor as a function of these quantities.

\subsubsection{Perfect fluid stress tensor}\label{sec:perfstr}

The dynamics of a fluid is completely specified once the stress tensor and charge currents are given as functions of $\tloc$ 
and $u^\mu$. As we have explained in the introduction, fluid mechanics is an effective description at long distances (i.e, it is valid only when the fluid variables vary on distance scales that are large compared to the mean free path $l_\mathrm{mfp}$). As a consequence it is natural to expand the stress tensor 
in powers of derivatives. In this subsection we briefly review the leading (i.e.\ zeroth) order terms in this expansion.

It is convenient to define a projection tensor
\begin{equation}\label{proj:eq}
  P^{\mu\nu} = g^{\mu\nu} + u^\mu u^\nu.
\end{equation}
$P^{\mu\nu}$ projects vectors onto the $(d-1)$ dimensional submanifold orthogonal to $u^\mu$. In other words, $P^{\mu\nu}$ may be thought of as a projector onto spatial coordinates in the rest frame of the fluid. In a similar fashion, $- u^\mu u^\nu$ projects vectors onto the time direction in the fluid rest frame.

To zeroth order in the derivative expansion, Lorentz invariance and the correct static limit uniquely determine the stress tensor, charge and the entropy currents in terms of the thermodynamic variables. We have
\begin{equation}\label{currents:eq}
\begin{split}
  T^{\mu\nu}_\mathrm{perfect}& = \rho u^\mu u^\nu + \ploc P^{\mu\nu}, \\
  (J^\mu_S)_\mathrm{perfect}&=s u^\mu,
\end{split}
\end{equation}
where all thermodynamic quantities are measured in the local rest frame of the fluid, so that they are Lorentz scalars. It is not difficult to verify that in this zero-derivative (or perfect fluid) approximation, the entropy current is conserved. Entropy production (associated with dissipation) occurs only at the first subleading order in the derivative expansion, as we will see in the next subsection.

\subsubsection{Dissipation}\label{sec:visc}

Now, we proceed to examine the first subleading order in the derivative expansion. In the first subleading order, Lorentz invariance and the physical requirement that entropy be non-decreasing determine the form of the stress tensor and the current to be (see, for example, \S\S14.1 of \cite{Andersson:2006nr})
\begin{equation}\label{extraTvisc:eq}
\begin{split}
  T^{\mu\nu}_\mathrm{dissipative} &= -\zeta \vartheta P^{\mu\nu} -
       2\eta\sigma^{\mu\nu} + q^\mu u^\nu + u^\mu q^\nu,\\
  (J^\mu_S)_\mathrm{dissipative} &= \frac{q^\mu }
                                         {\tloc}\,.
\end{split}
\end{equation}
where
\begin{equation}\label{fluidtensors:eq}
\begin{split}
  a^\mu &= u^\nu \nabla_\nu u^\mu, \\
  \vartheta &= \nabla_\mu u^\mu, \\
  \sigma^{\mu\nu} &= \half \prn{P^{\mu\lambda} \nabla_\lambda u^\nu
                   + P^{\nu\lambda} \nabla_\lambda u^\mu}
                   - \frac{1}{d-1} \vartheta P^{\mu\nu}, \\
 q^\mu &= -\kappa P^{\mu\nu} (\p_\nu\tloc + a_\nu\tloc)\,, \\
\end{split}
\end{equation}
are the acceleration, expansion, shear tensor and heat flux
respectively.

These equations define a set of new fluid dynamical parameters in addition to those of the previous subsection: $\zeta$ is the bulk viscosity, $\eta$ is the shear viscosity and $\kappa$ is the thermal conductivity. 
Fourier's law of heat conduction $\vec{q} = -\kappa \vec{\nabla} \tloc$ has been relativistically modified to
\begin{equation}\label{heatcond:eq}
  q^\mu = -\kappa P^{\mu\nu} (\p_\nu\tloc + a_\nu\tloc)\,,
\end{equation}
with an extra term that is related to the redshifting of the temperature. 

At this order in the derivative expansion, the entropy current is no longer conserved; however, it may be checked \cite{Andersson:2006nr} that
\begin{equation}\label{increase:eq}
 \tloc\nabla_\mu J^\mu_S = \frac{q^\mu q_\mu}{\kappa \tloc}
 + \zeta \vartheta ^2 + 2 \eta \sigma_{\mu \nu} \sigma^{\mu \nu}.
\end{equation}
As $q^\mu$ 
and $\sigma^{\mu \nu}$ are all spacelike
vectors and tensors, the RHS of \eqref{increase:eq} is positive provided
$\eta, \zeta, \kappa$ 
are positive parameters, a condition we further assume. This establishes that (even locally) entropy can only be produced but never destroyed. In equilibrium, $\nabla_\mu J^\mu_S$ must vanish. It follows that, $q^\mu$, 
$\vartheta$ and $\sigma^{\mu \nu}$ each individually vanish in equilibrium.

For fluids with gravity duals, the shear viscosity is given by $\eta=\frac{s}{4\pi}$ \cite{Son:2007vk}. We can estimate the thermalisation length of the fluid by comparing coefficients at different orders in the derivative expansion
\begin{equation}\label{mfp:eq}
  \mfp \sim \frac{\eta}{\rho} = \frac{s}{4\pi\rho}.
\end{equation}
This length scale may plausibly be identified with the thermalisation length scale of the fluid. This may be demonstrated within the kinetic theory, where $\mfp$ is simply the mean free path of colliding molecules, but is expected to apply to more generally to any fluid with short range interactions.

With the equation of state \eqref{bbeos:eq}, this is given by
\begin{equation}\label{bbmfp:eq}
  \mfp \sim \frac{\tloc^d}{\pi(d\tloc^{d+1}+\tc^{d+1})}.
\end{equation}
As we will be restricting attention to temperatures close to $\tc$, we have $\mfp \sim 1/\tc$.

\subsection{Surfaces}\label{sec:surface}

The plasma ball configurations we consider have a domain wall separating a bubble of the deconfined phase from the confined phase. As the density, pressure, etc.\ of the deconfined phase are a factor of $N^2$ larger than the confined phase, we can treat the confined phase as the vacuum and the domain wall as a surface bounding the deconfined fluid.

At surfaces, the density of the fluid changes too rapidly to be described by fluid mechanics. However, provided that we look at length scales much larger than the thickness of the surface, we can replace this region by a delta function localised piece of the stress tensor.

At these length scales, this stress tensor will depend on the direction of the surface, with dependence on its curvature being suppressed.

In general, introducing a surface energy density $\sigma_E$, a surface entropy density $\sigma_S$ and a surface tension $\sigma$, 
considerations similar to those leading to \eqref{inttherm:eq} lead to
\begin{equation*}
  \begin{split}
    \sigma_E &= \sigma + \tloc \sigma_S ,
    \\
    \dr\sigma &= -\sigma_S\, \dr\tloc .
  \end{split}
\end{equation*}
However, the surface tension was only computed at $\tloc=\tc$ 
in \cite{Aharony:2005bm}, so we will have to ignore its temperature 
dependence. As we can see above, this is equivalent to setting $\sigma_S
=0$ and $\sigma_E=\sigma$.

Let's describe the location of the surface by a function $f(x)$ that is positive inside the fluid and has a first order zero on the surface:
\begin{equation}\label{fluidsurf:eq}
  T^{\mu\nu} = \theta(f)T^{\mu\nu}_\text{fluid} + \delta(f)T^{\mu\nu}_\text{surface}.
\end{equation}
At large length scales, as mentioned above, $T^{\mu\nu}_\text{surface}$ will only depend on the first derivative of $f$ and no higher derivatives.

If we demand invariance under reparameterisations of the function $f(x) \to g(x)f(x)$, where $g(x) > 0$, and that the surface moves at the velocity of the fluid
\begin{equation}\label{surfvel:eq}
  \left. u^\mu \p_\mu f \right\vert_{f=0} =0,
\end{equation}
the surface stress tensor is (see \S2.3 of \cite{Lahiri:2007ae})
\begin{equation}\label{surfstressgen:eq}
  T^{\mu\nu}_\text{surface} = \sqrt{\p f \cdt \p f} \brk{\sigma_E u^\mu u^\nu -\sigma(g^{\mu\nu}- n^\mu n^\nu + u^\mu u^\nu)},
\end{equation}
where $n_\mu = -\p_\mu f / \sqrt{\p f \cdt \p f}$ is the normal to the surface. Note that $(\p_\mu f) T^{\mu\nu}_\text{surface}=0$. If we take the surface tension to be constant, as above, we get
\begin{equation}\label{surfstress:eq}
  T^{\mu\nu}_\text{surface} = -\sigma h^{\mu\nu} \sqrt{\p f \cdt \p f},
\end{equation}
where $h_{\mu\nu} = g_{\mu\nu} - n_\mu n_\nu$ is the induced metric of the surface. The factor of $\sqrt{\p f \cdt \p f}$ also has a simple interpretation: suppose we use a coordinate system where $f$ is one of the coordinates. Then
\begin{equation}\label{surfmeasure:eq}
  \sqrt{\p f \cdt \p f} = \sqrt{g^{ff}} = \sqrt{\frac{\det h}{\det g}}\,,
\end{equation}
which provides the correct change of integration measure for localisation to the surface. If we used some other coordinates, there'd be an extra Jacobian factor.

We have
\begin{equation}\label{surfeom:eq}
  \nabla_\mu T^{\mu\nu} = \theta(f)\nabla_\mu T^{\mu\nu}_\text{fluid} + \delta(f)(\p_\mu f) T^{\mu\nu}_\text{fluid} + \delta(f)\nabla_\mu T^{\mu\nu}_\text{surface}.
\end{equation}
So, in addition to \eqref{Epconsv:eq}, we have the boundary conditions
\begin{equation}\label{surfbc:eq}
  \left.\phantom{\half}(\p_\mu f) T^{\mu\nu}_\text{fluid} + \nabla_\mu T^{\mu\nu}_\text{surface}\right\vert_{f=0}=0.
\end{equation}
Also, when we take the surface tension to be constant:
\begin{equation}\label{extcurvsurf:eq}
  \nabla_\mu T^{\mu\nu}_\text{surface} = \sigma \brk{\frac{\square f}{(\p f \cdt \p f)^{1/2}} - \frac{(\p^\mu\! f)
 (\p^\lambda\! f)\nabla_\mu\p_\lambda f }{(\p f \cdt \p f)^{3/2}}} \p^\nu\! f = -\sigma \,\Theta \, \p^\nu\! f,
\end{equation}
where $\Theta$ is the trace of the extrinsic curvature of the surface, as seen from outside the fluid (see Appendix \ref{sec:extrinsic}).

If we have several disconnected surfaces, it is convenient to make the separation $f=\prod_i f_i$. As the surfaces are disconnected, the zero sets of the $f_i$ do not intersect. Also, the $f_i$ are all positive inside the fluid. Therefore, whenever one of the $f_i$ is negative or zero, all the others are positive. Luckily, \eqref{surfeom:eq} splits nicely
\begin{equation*}
  \nabla_\mu T^{\mu\nu} = \prod_i \theta\prn{f_i}\nabla_\mu T^{\mu\nu}_\text{fluid} + \sum_i \delta(f_i)\brk{(\p_\mu f_i) T^{\mu\nu}_\text{fluid} + \nabla_\mu T^{\mu\nu}_\text{surface}(f_i)}.
\end{equation*}

From the form of the gravity solution, we would expect $\sigma_E/\rho$ to be similar to the thickness of the surface. We can estimate it using

\begin{equation}\label{thick:eq}
  \xi = \frac{\sigma}{\rc} = \frac{\sigma}{(d+1)\eos\tc^d}\,.
\end{equation}
In general, it will be of order $N^0$ and is similar to the surface thickness and $\mfp$ (if $8\pi$ can be considered similar to 1).

For the domain wall of \cite{Aharony:2005bm} in $d=2+1$ dimensions, the thickness and surface tension are approximately $6\times \frac{1}{2\pi \tc}$ and $\sigma=2.0 \times \frac{\rc}{\tc}$ respectively. This gives $\xi = \frac{2.0}{\tc}$, which is pretty close to the thickness.

In $d=3+1$, the domain wall of \cite{Aharony:2005bm} has thickness and surface tension approximately equal to $5\times \frac{1}{2\pi \tc}$ and $\sigma=1.7 \times \frac{\rc}{\tc}$ respectively. This gives $\xi = \frac{1.7}{\tc}$, which is also pretty close to the thickness.

For our purposes, it is more convenient to talk about the length scale
\begin{equation}\label{xipr:eq}
  \xi' = (d+1)\xi = \frac{\sigma}{\rz} = \frac{\sigma}{\eos\tc^d}\,. \end{equation}

\subsection{Equilibrium configurations}\label{sec:rigidrot}

In this subsection, we will specialise the general discussion above to the construction of equilibrium configurations of fluids with surfaces. We will also derive a simple approach to studying the thermodynamic properties of these configurations.

\subsubsection{Solutions for the interior}\label{sec:rotint}

We want to find solutions of \eqref{Epconsv:eq} that are independent of time, which means we need to set \eqref{increase:eq} to zero. This means we need velocity configurations that have zero expansion and shear. In general, this would be a combination of a uniform boost and rigid rotation. We can always boost to a frame where the centre of rotation is static and the rotation lies in the Cartan directions of the rotation group. This gives
\begin{equation}\label{rigidrot:eq}
  u = \gamma(\p_t + \omega_a l_a),
\end{equation}
where $\omega_a$ are the angular velocities and $l_a$ are a set of commuting rotational Killing vectors. The important feature is that the velocity is a normalisation factor times a Killing vector (see \S2.2 of \cite{Caldarelli:2008mv}):
\begin{equation}\label{eqvel:eq}
  u^\mu = \gamma K^\mu, \qquad
  \gamma^2 K^\mu K_\mu = -1, \qquad
  \nabla_{(\mu} K_{\nu)} = 0.
\end{equation}
One can deduce that
\begin{equation*}
  \vartheta = \sigma^{\mu\nu}=0, \qquad
  u^\mu \p_\mu \gamma = 0, \qquad
  a_\mu = -\frac{\p_\mu \gamma}{\gamma}\,.
\end{equation*}
Which leads to
\begin{equation*}
  q^\mu = -\kappa \gamma P^{\mu\nu} \p_\nu \brk{\frac{\tloc}{\gamma}}.
\end{equation*}
One can also show that
\begin{equation*}
  \nabla_\mu T^{\mu\nu}_\mathrm{perfect} =
    \gamma \prn{s P^{\nu\mu}
        + 
        \tloc \pdiff{s}{\tloc} 
         u^\nu u^\mu}
       \p_\mu \brk{\frac{\tloc}{\gamma}}
\end{equation*}
So the velocity configuration \eqref{rigidrot:eq} will be an equilibrium solution to the equations of motion provided that
\begin{equation}\label{rotsol:eq}
  \frac{\tloc}{\gamma} = T = \text{constant} .
\end{equation}
Using the equation of state and \eqref{inttherm:eq}, this determines all of the intensive thermodynamic quantities in the fluid.

\subsubsection{Solutions for surfaces}\label{sec:rotsurf}

The fluid configurations described in the previous subsection have $T^{\mu\nu}_\mathrm{dissipative}=0$. Therefore
\begin{equation*}
  (\p_\mu f) T^{\mu\nu}_\text{fluid} = (\p_\mu f) T^{\mu\nu}_\text{perfect} = \ploc \p^\nu\! f.
\end{equation*}
This means that \eqref{surfbc:eq} and \eqref{extcurvsurf:eq} reduce to
\begin{equation}\label{rotsurf:eq}
  \ploc|_{f=0} = \sigma \Theta.
\end{equation}
As the pressure is determined by \eqref{rotsol:eq}, this provides a differential equation that determines allowed positions of surfaces. Demanding that the surface has no conical singularities turns out to provide enough boundary conditions to determine the position of the surface completely (up to discrete choices) in terms of the parameters $\Omega_a$, $T$.

\subsubsection{Thermodynamics of solutions}\label{sec:rottherm}

We compute the extensive thermodynamic properties of these solutions by integrating the time components of the corresponding currents (noting that the current associated with a Killing vector $\zeta^\mu$ is $J^\mu_\zeta = T^{\mu\nu}\zeta_\nu$):
\begin{equation}\label{noetherch:eq}
 \begin{split}
  Q_X &= \int\!\dr V J^0_X.
 \end{split}
\end{equation}
We are assuming that the space-time in consideration is static, so it can be foliated by space-like surfaces $t=\text{constant}$ with normal $\p_t$. In fact, here we will only consider fluids in flat space.

In particular, also noting that for equilibrium configurations $\p^0\!f=0$,
\begin{equation}\label{killingcharge:eq}
  Q_\zeta = \int\!\dr V \theta(f)\brk{(\rho+\ploc)\gamma^2 K^0 K\cdt\zeta
   + \ploc \zeta^0 } - \int\!\dr V \delta(f) \sqrt{\p f\cdt\p f} \sigma \zeta^0.
\end{equation}
Noting that $K^0=(\p_t)^0=1$ and $l_a^0=0$, this gives
\begin{equation}\label{thermcharge:eq}
  \begin{aligned}
    E &= -Q_{\p_t}& &= -\int\!\dr V \theta(f)\brk{
           (\rho+\ploc)\gamma^2 K\cdt\p_t +\ploc}
       + \int\!\dr V \delta(f) \sqrt{\p f\cdt\p f} \sigma, \\
    L_a &= Q_{l_a}& &= \int\!\dr V \theta(f)\brk{
          (\rho+\ploc)\gamma^2 K\cdt l_a},\\
    S &= Q_S& &= \int\!\dr V \theta(f)\brk{\gamma s}.
  \end{aligned}
\end{equation}

From these quantities, we can compute overall angular velocities $\omega_a$ and 
temperature $T$ 
thermodynamically
\begin{equation}\label{chpotdef:eq}
  \dr E = \omega_a \,\dr L_a + T \,\dr S.
\end{equation}
Note that these quantities are different from the local thermodynamic
properties of the fluid in its rest frame. In particular, the local
temperature, $\tloc$, only knows about the thermal energy of the plasma,
whereas the overall temperature, $T$, also knows about its kinetic energy.
\emph{A priori}, it may not seems that these quantities have to be the same as $\omega_a$, $T$ 
from \eqref{rigidrot:eq} and \eqref{rotsol:eq}. However, we can show that they are the same by checking that \eqref{chpotdef:eq} holds with $\omega_a$, $T$ 
taken from \eqref{rigidrot:eq} and \eqref{rotsol:eq}. In practice, it is easier to verify the equivalent statement
\begin{equation}\label{chpotcheck:eq}
  \dr(E -\omega_a L_a - T S )
  = - L_a \,\dr \omega_a - S \,\dr T .
\end{equation}

First, making use of \eqref{inttherm:eq}, we see that
\begin{equation}\label{thermpot:eq}
  E -\omega_a L_a - T S 
   = -Q_K - T Q_S 
   = - \int\!\dr V \theta(f) \ploc + \int\!\dr V \delta(f) \sqrt{\p f\cdt\p f} \sigma.
\end{equation}
Note that the second integral is simply $\sigma$ times the surface area: as we saw in \eqref{surfmeasure:eq} the factor of $\sqrt{\p f\cdt\p f}$ provides the correct change of measure for the delta function to localise the integral to the surface.

Consider an infinitesimal change of $\omega_a$, $T$. 
We have
\begin{equation*}
\begin{split}
  \dr \ploc &= s\,\dr(\gamma T) 
  = \frac{\rho+\ploc}{\gamma}\,\dr\gamma + \gamma s\,\dr T .\\
  \gamma^{-3}\,\dr\gamma &= K\cdt\dr K = K\cdt l_a \,\dr\omega_a.
\end{split}
\end{equation*}
From this, we see that \eqref{chpotcheck:eq} is satisfied by the contributions from the interior. As the right hand side of \eqref{chpotcheck:eq} has no contributions from the surface, we need to check that the surface contributions of the variation of \eqref{thermpot:eq} cancel.

The change in the surface area can be written as
\begin{equation*}
  \dr\mathcal{A} = \oint\dr A\, \vec{n}\cdt\vec{w},
\end{equation*}
where the integral is performed over the union of the initial and final surfaces, $\vec{n}$ is a unit normal vector pointing into the initial fluid and out of the final fluid and $\vec{w}$ is some vector field that is equal to the outward pointing normal at both the initial and final surfaces. By Gauss' theorem, this can be written as
\begin{equation*}
  \dr\mathcal{A} = \int\!\dr V\, \nabla\cdt\vec{w},
\end{equation*}
with the integral performed over the region between the two surfaces. The volume element can be written as $\int\!\dr V = \int\!\dr A\,(\vec{n}\cdt\Delta x)$, with $\vec{n}$ pointing outwards. As the volume element is already infinitesimal, we can replace $\vec{w}$ with the vector field described in \eqref{normoffsurf:eq}, as the difference would be infinitesimal, i.e.\ $\nabla\cdt\vec{w} \to \Theta$. Also, as $f=0$ on the initial surface, and $f+\dr f=0$ on the final surface ($\dr f$ refers to the change in $f$ due to the change in $\Omega_a$, $T$)
, we have
\begin{equation*}
\begin{split}
  \p_\mu f \Delta x^\mu + \pdiff{f}{\Omega_a}\dr\Omega_a + \pdiff{f}{T}\dr T 
  = 0,\\
  \implies \vec{n}\cdt\Delta x = \frac{\dr f}{\sqrt{\p f\cdt\p f}}.
\end{split}
\end{equation*}
Therefore
\begin{equation*}
  \dr\mathcal{A} = \int\!\dr V \delta(f) \Theta \,\dr f.
\end{equation*}
So, we can write the surface contribution to the variation of \eqref{thermpot:eq} as
\begin{equation*}
  \dr(E -\omega_a L_a - T S 
  )_{\text{surface}}
  =-\int\!\dr V \delta(f) \,\ploc \,\dr f + \int\!\dr V \delta(f)\,\sigma \Theta \,\dr f,
\end{equation*}
which vanishes due to \eqref{rotsurfbc:eq}.

The thermodynamics of the solution can be summarised by defining a grand partition function
\begin{equation}\label{gpf:eq}
  \gpf = \Tr\exp\prn{-\frac{E -\omega_a L_a  
  }{T}}.
\end{equation}
In the thermodynamic limit,
\begin{equation}\label{gpftherm:eq}
  \begin{split}
    -T\ln\gpf &= E -\omega_a L_a - T S 
    , \\
    \dr(T\ln\gpf) &= L_a \,\dr \omega_a + S \,\dr T .
  \end{split}
\end{equation}
We have seen that
\begin{equation}\label{gpfrot:eq}
  T\ln\gpf = \int_{f>0}\!\!\!\dr V \,\ploc - \int_{f=0}\!\!\!\dr A \,\sigma
\end{equation}
and the $\omega_a$, $T$ 
are the same as those given by \eqref{rigidrot:eq} and \eqref{rotsol:eq}.

\subsubsection{Validity}\label{sec:validity}

We are making several approximations in this paper. First of all, the Navier-Stokes equations are merely a long wavelength approximation to the full dynamics of the gauge theory. This will be a good approximation provided that the length scale of variation of $\tloc$ and $u^\mu$ is small compared to the thermalisation scale of the fluid \eqref{mfp:eq}.

Second, we have treated the surface of the plasma as sharp; in
reality this surface has a thickness of order $\xi$ (see \eqref{thick:eq}).
Consequently, our treatment of the surface is valid only when its
curvature is small compared to
$1/\xi \sim 1/\xi'$ (higher derivative contributions to the surface stress
tensor, which we have ignored in our treatment, would become
important if this were not the case); further we must also require
that only a small fraction of the fluid should reside in surfaces. This boils down to demanding that all sizes are much larger than $\xi'$.

Thirdly, we have ignored the fact that the surface tension is a
function of the fluid temperature at the surface, and simply set
$\sigma=\sigma(\tc)$. This is valid provided that $\tloc/\tc \approx
1$ at all surfaces. When this is the case, the pressure will be small compared to $\rc$ (see \eqref{bbeos:eq}). Then, \eqref{rotsurfbc:eq} tells us that the extrinsic curvature of the surface must be small compared to $1/\xi$, which is the same as the previous condition.

We can estimate the scale over which thermodynamic quantities
vary as the distance over which the fractional change in the temperature is one. As the temperature is proportional
to $\gamma$, we should demand (schematically)
\begin{equation*}
  \frac{1}{\nrm{\nabla\ln\gamma}} \sim \frac{1-v^2}{\nrm{\omega v}}
  \gg \mfp\,.
\end{equation*}
At temperatures close to $\tc$, where our other approximations are valid, we have $\mfp \sim \xi'$. Therefore, we require that that
\begin{equation*}
  \frac{1-v^2}{\nrm{\omega v}} \gg \xi'.
\end{equation*}
This will be true if the speed of the fluid is much less than the speed of light and the angular velocities are much less that $1/\xi'$.

\subsubsection{Dimensionless variables}\label{sec:units}

It is convenient to rescale the variables as follows. First we rescale all lengths and times by $\xi'$ (i.e.\ working in units where $\xi'=1$)
\begin{equation}\label{distunits:eq}
  x = \xi' \tilde{x},  \qquad
  \omega_a = \frac{\tw_a}{\xi'}\,.
\end{equation}
We measure temperature in units of $\tc$ and further rescale extensive thermodynamic quantities by $\eos\tc^d$.
\begin{equation}\label{thermunits:eq}
  \begin{aligned}
    T &= \tc \tT, \quad&
    \ln\gpf &= (\eos\tc^{d-1} \xi'^{d-1}) \ln\tgpf, \quad&
    E &= (\eos\tc^{d} \xi'^{d-1}) \tE, \\
    S &= (\eos\tc^{d-1} \xi'^{d-1}) \tS, &
    L &= (\eos\tc^{d} \xi'^{d}) \tL.
  \end{aligned}
\end{equation}
Then, \eqref{rotsurfrepeat:eq} becomes
\begin{equation}\label{rotsurfbc:eq}
  \brk{(\gamma\tT)^{d+1} - 1}_\text{surface} = \widetilde{\Theta},
\end{equation}
and \eqref{rotsoltherm:eq} becomes
\begin{equation}\label{rotthermscaled:eq}
  \begin{split}
    \tT\ln\tgpf &= \int_{f>0}\!\!\!\dr \widetilde{V} \brk{(\gamma\tT)^{d+1} - 1 } - \int_{f=0}\!\!\!\dr \widetilde{A} , \\
    -\tT\ln\tgpf &= \tE -\tw_a \tL_a -\tT \tS,\\ 
    \dr(\tT\ln\tgpf) &= \tL_a \,\dr \tw_a + \tS \,\dr\tT.
  \end{split}
\end{equation}

From now on, we will suppress all tildes and work entirely with the new variables.

\subsubsection{Summary}\label{sec:rigidsummary}

We can summarise the construction of equilibrium solutions as follows:

The fluid velocity is given by
\begin{equation}\label{rotvel:eq}
  \begin{split}
    u^\mu = \gamma K^\mu, \qquad \text{where}\quad
     K &= \p_t + \omega_a \p_{\phi_a},  \\
     \gamma &= (-K^\mu K_\mu)^{-\half},
  \end{split}
\end{equation}
with $\phi_a$ being a set of angular coordinates such that $\phi_a \to \phi_a + c_a$ are a set of commuting isometries and $\omega_a$ the angular velocities.

The thermodynamic properties of the fluid are specified by
\begin{equation}\label{rotfluidtherm:eq}
  \tloc = \gamma T.
\end{equation}
All other local thermodynamic properties can be computed from the equation of state \eqref{bbeos:eq} and the relations \eqref{inttherm:eq}.

The position of the surface is specified by a function $f$ that is positive inside the fluid and negative outside. It is determined by
\begin{equation}\label{rotsurfrepeat:eq}
  \brk{(\gamma T)^{d+1} - 1}_\text{surface} = \Theta,
\end{equation}
with $\Theta$ given by \eqref{trextrsurf:eq}, and the condition that the surface is closed without conical singularities.

The overall thermodynamic properties of the solution can be computed from
\begin{equation}\label{rotsoltherm:eq}
  \begin{split}
    T\ln\gpf &= \int_{f>0}\!\!\!\dr V \brk{(\gamma T)^{d+1} - 1} - \int_{f=0}\!\!\!\dr A \,, \\
    -T\ln\gpf &= E -\omega_a L_a - T S,\\ 
    \dr(T\ln\gpf) &= L_a \,\dr \omega_a + S \,\dr T.
  \end{split}
\end{equation}
%

\section{Black objects in arbitrary dimensions}\label{sec:arbdim}

As mentioned in the introduction we study the horizon topologies of black objects in SSAdS$_{d+2}$ through
the dual fluid configurations which solve the $d$ dimensional relativistic Navier-Stokes equation.
In \cite{Lahiri:2007ae} exact disc ($B^2$) like plasma configurations were obtained in 2+1 dimensions.
In one higher dimension (i.e. in 3+1 dimensions) we can expect the existence of the solution $B^2 \times \R^1$,
which is topologically a cylinder. Now we can bend the cylinder into a ring. The fact that such a ring
solution exists were shown (numerically) in \cite{Lahiri:2007ae} and
its thermodynamics were explored in \cite{Bhardwaj:2008if}. This ring solution can be constructed perturbatively from the cylinder
(a suitable expansion parameter being the ratio $\epsilon = \frac{R}{\mathcal{L}}$,
$R$ being the radius of the cylinder and $\mathcal{L}$ being the
distance of the cylinder from the origin). The leading order solution, called the \emph{thin ring},
(correct up to $\CO(\epsilon)$) was obtained analytically in \cite{Bhardwaj:2008if}.

This method can be used to construct rings and other exotic plasma configurations in higher dimensions. For example in one more
dimension (i.e. 4+1 dimensions) we can have a cylinder solution with topology $B^3 \times \R^1$ ($B^3$ being the ball type solution
in $3+1$ dimension) then we can bend this cylinder to form a ring. In this way we would be able to obtain ring
type solutions from the
plasmaball configuration in one lower dimension. Further besides the ring we can also construct other configurations by a similar method.
Again for concreteness let us consider the example of $4+1$ dimension where besides the $B^3 \times \R^1$ topology we can also have a topology
$B^2 \times \R^2$ constructed out of the $B^2$ solution in $2+1$ dimensions (two dimensions lower). Now we can bend the $B^2 \times \R^2$
configuration along two directions in the $\R^2$ to form $B^2 \times S^1 \times S^1$. Another way to think of it is that
in $3+1$ dimensions the $B^2 \times S^1$ topology existed. So (just like the construction of the cylinder) in $4+1$ dimension
we have a topology $B^2 \times S^1 \times \R^1$. Now we can bend this cylinder along the $\R^1$ to form
the topology $B^2 \times S^1 \times S^1$.
Thus in $d$ space-time dimensions we  can use this method to study a topology
\begin{equation}\label{topo:eq}
 B^{(d-1-n)} \times T^n = B^{(d-1-n)} \times \underbrace{ S^1 \times S^1 \dots  \dots S^1}_{\text{n times}}.
\end{equation}
Note that there can be solutions with other topologies in $d$ dimensions which we will not be able to capture by our
method, therefore we shall look at these topologies only.
Here $n$ cannot be arbitrary. Naively we would expect $n \leq d-3$ so that the ball is at least a $B^2$
\footnote{ If the ball is a $B^1$ then we would be considering hollow solutions; however we are unable to capture the
description of those solutions by our method. In addition, such hollow solutions might not exist in all dimension.
In $2+1$ dimensions the existence of such hollow solutions (the annular ring) was reported in \cite{Lahiri:2007ae}; while
there is strong evidence (see \cite{Lahiri:2007ae, Bhardwaj:2008if}) to suggest that such hollow solutions
do not exist in $3+1$ dimensions.}. However there is a stronger
bound on $n$. We recall the fact that for the ring (or more exotic objects as above) to exist we must have non-zero
angular momentum in the plane of the $S^1$(s). However in $d$ dimensions we can independently turn
on angular momentum only along $\frac{d-1}{2}$ ($\frac{d-2}{2}$) directions for odd (even) $d$. This is because the
group of spatial rotations in $d$ dimensions is $SO(d-1)$ which has rank $\frac{d-1}{2}$ ($\frac{d-2}{2}$) for odd (even) $d$.
Thus $n$ should not  be more than the number of Cartans of the (spatial) rotation group in a particular dimension.
Hence $n$ should satisfy
\begin{equation}
 \begin{split}
  n &= 0 , \quad \text{for } d=3.\\
  n &\leq \frac{d-1}{2}, \quad \text{for odd } d \text{ greater than } 3.\\
  n &\leq \frac{d-2}{2}, \quad \text{for even d}.
 \end{split}
\end{equation}

Here note that $d=3$ is a special case because for this dimension $n \leq d-3$ is a stronger bound than $n \leq \frac{d-1}{2}$.
Also note that a new topology of the plasma configuration is obtained for every odd d $(\geq 5)$. Thus in an even $d$ we only
have the solutions that existed in $d-1$ dimensions.

Here we shall analytically show (in an approximation analogous to the thin ring approximation)  that these topologies exist as
plasma configurations which are solutions of the relativistic Navier-Stokes equations expressed in the form \eqref{Epconsv:eq}.
The most suitable coordinate system for studying the general topology \eqref{topo:eq} is the one in which we choose $\{r_a, \phi_a\}$
with $a=1,\dots n$ as the coordinates on the $n$ planes containing the $n$ $S^1$s and in the rest of the space
(which exists exists when $d > 2n+1$) we choose spherical polar coordinates $\{r,\theta_1, \dots , \theta_{(d-2n-2)}\}$.
The coordinates $\phi_a$ and $\theta_j$ are angular coordinates; the coordinates $\{\theta_j\}$ with $j=1, \dots,(d-2n-2)$
may be taken to be the coordinates on a unit sphere in a $d-2n-1$ dimensional space.
The angles $\theta_j$ will be absent if $d=2n+2$. The range of all the
radial coordinates $\{r,r_a\}$ are as usual $\left[0,\infty \right)$
while the angles $\phi_a \in \left[0 ,2 \pi \right)$. Besides these spatial coordinate
we denote the time coordinate by $t$.

The metric is given by
\begin{equation}\label{genmet:eq}
 \dr s^2 = -\dr t^2 + \sum_{a=1}^{n} r_a^2 \dr\phi_a^2
           + \sum_{a=1}^{n} \dr r_a^2 + \dr r^2
           + r^{(d-2n-2)} \dr\Omega_{(d-2n-2)}^2(\theta_1,\theta_2, \dots, \theta_{d-2n-2})
\end{equation}
where $\dr\Omega_{(d-2n-2)}^2$ is the round metric on a unit sphere in $d-2n-1$ dimensions.
In such a space we consider the fluid surface to be given by

\begin{equation}
 f \equiv h(r_1,r_2, \dots,r_n) - r = 0.
\end{equation}

For $n=\frac{d-1}{2}$ we have a  special case because then there is no $r$ coordinate.
In this case we consider $f \equiv h(r_2,\dots,r_n ) - r_1 = 0$ as the
equation of the surface.

The velocity is given by $u^{\mu}=\gamma (1,\omega_1,\dots,\omega_n,0,0,\dots,0)$, with
$\gamma = (1-\sum_{a=1}^n \omega_a^2 r_a^2)^{\frac{1}{2}}$ being the normalization factor.
Note that here we are not considering any angular velocity in the $\theta_i$ directions.
The topologies that we explore could also be spinning
along the $\theta_i$ directions. However, if that were the case the zeroth order solution would not be simply a round ball times a plane and we would lose analytic control. It might be possible to turn on infinitesimal angular velocities in these directions, but we will not consider that here.

The equation for h($r_a$) that follows from \eqref{rotsurfbc:eq} is given by
\begin{equation}\label{genmaineq}
\begin{split}
 \frac{T^{d+1}}{(1-\sum_{a=1}^n \omega_a^2 r_a^2)^{(\frac{d+1}{2})}}-1&
 = \frac{1}{(1+\sum_{a=1}^n(\partial_{r_a}  h) (\partial_{r_a} h))^{\frac{3}{2}}}
 \left( \left(\frac{m}{h}-\sum_{a=1}^n \frac{\partial_{r_a} h}{r_a}
 - \sum_{a=1}^n\partial_{r_a} \partial_{r_a} h \right) \right. \\
 &\quad \quad \quad \left( 1 + \sum_{a=1}^n {(\partial_{r_a} h)( \partial_{r_a} h)} \right)
 \left. +\sum_{a,b=1}^n (\partial_{r_a} h) (\partial_{r_b} h) (\partial_{r_a} \partial_{r_b} h) \right).
\end{split}
\end{equation}

where $m = d-2n-2$.
In order to obtain \eqref{genmaineq} from \eqref{rotsurfbc:eq} we set the typical length scale of the
problem $\xi'$ to 1 by suitable choice of units and we are measuring temperature in units of $\tc$, as described in \S\ref{sec:units}.
For $n=\frac{d-1}{2}$ the sum starts running from 2 (instead of 1) in the above equation.
Now we separately consider the space of $\{r,r_a\}$ (only the first 'quadrant' of this
space is physical because $\{r,r_a\} \in [0,\infty)$).
We now shift the origin to a new point $\overrightarrow{\mathcal{L}}$
such that it is given by
\begin{equation}\label{projcoord:eq}
 \overrightarrow{\mathcal{L}} = (0,\mathcal{L} P_1,\mathcal{L} P_2,\dots,\mathcal{L} P_n)
\end{equation}
$P_a$s are projectors along various $r_a$ directions, so that $\sum P_a^2 = 1$ and
$\mathcal{L}$ is the magnitude of the vector $\overrightarrow{\mathcal{L}}$. Again in the special case of
$n=\frac{d-1}{2}$ we take $\overrightarrow{\mathcal{L}} = (\mathcal{L} P_1,\mathcal{L} P_2,\dots,\mathcal{L} P_n)$.
Let $\{x_a\}$ be the new shifted coordinates such that
\begin{equation}\label{shift:eq}
 r_a = \mathcal{L} P_a + x_a .
\end{equation}
As is apparent the coordinate $r$ (if it exists) remains unchanged by this coordinate change.

Now we perform the following scaling
\begin{equation}\label{scalegen:eq}
 \omega_a = \epsilon w_a; \quad
 \mathcal{L} = \frac{\ell_0}{\epsilon}; \quad
 h = y(\{x_a\}).
\end{equation}
Then \eqref{genmaineq} at leading order in $\epsilon$ (which is $\epsilon^0$)
reduces to the equation
\begin{equation}\label{order0eq}
\begin{split}
\frac{T^{d+1}}{\left( 1- \sum_{a} (w_a \ell_0 P_a)^2 \right)^{\left( \frac{d+1}{2} \right)}}-1
= &\frac{1}{(1+\sum_{a=1}^n{\partial_{x_a}  y \partial_{x_a} y})^{\frac{3}{2}}}
\left( \left(\frac{m}{y} - \sum_{a=1}^n\partial_{x_a} \partial_{x_a} y \right)
\right. \\
& \left(1+\sum_{a=1}^n{\partial_{x_a}  y \partial_{x_a} y}\right)
\left.+\sum_{a,b=1}^n (\partial_{x_a} y) (\partial_{x_b} y) (\partial_{x_s} \partial_{x_b} y) \right).
\end{split}
\end{equation}
Now \eqref{order0eq} is satisfied by the function
\begin{equation}\label{gensol}
 y(\{x_a\}) = \left( R^2 - \sum_{a=1}^n x_a^2 \right)^{\frac{1}{2}},
\end{equation}
provided the following equation is satisfied by the parameters
\begin{equation}\label{genres1}
 T^{d+1} = \left( \frac{(d-n-2)+R}{R} \right) \left( 1 - \sum_{a} (\ell_0 P_a w_a)^2\right)^{\left(\frac{d+1}{2}\right)}\\
\end{equation}
Also the equation \eqref{genmaineq} at $O(\epsilon)$ yields
\begin{equation}\label{order1eq}
 \frac{(d+1) T^{d+1} \sum_a (w_a^2 P_a \ell_0 x_a)}{\left(1-\sum_a (w_a P_a \ell_0)^2\right)^{\frac{d+3}{2}}}
+ \frac{1}{(1+\sum_{a=1}^n{\partial_{x_a}  y \partial_{x_a} y})^{\frac{1}{2}}} \sum_a \frac{\partial_{x_a} y}{\ell_0 P_a} =0.
\end{equation}
In the above equation if we substitute \eqref{gensol} and set the coefficients of $x_a$ to zero then
we get (after using \eqref{genres1})
\begin{equation}\label{genres2}
 w_a^2 = \frac{1}{\left(\ell_0 P_a \right)^2\left(((d-n-2)+R)(d+1)+n\right)}.
\end{equation}
Note that here we have $n+1$ equations for $n+2$ parameters $R$, $\ell_0$ and $n$ $P_a$s.
However  there is one constraint among the $P_a$s (namely $\sum P_a^2 =1$) which gives us the
correct number of equations for the parameters to be determined.

Again for the special case of $n=\frac{d-1}{2}$ the chief results \eqref{gensol}, \eqref{genres1} and \eqref{genres2}
remain unchanged. However we have $h = \mathcal{L} P_1 + y(\{x_i\})$.
As a result the 0th and 1st  order equations, \eqref{order0eq} and \eqref{order1eq}, are changed
respectively to
\begin{equation}\label{order0eqsp}
 \begin{split}
\frac{T^{d+1}}{\left( 1- \sum_{a} (w_a \ell_0 P_a)^2 \right)^{\left( \frac{d+1}{2} \right)}}-1
= &\frac{1}{(1+\sum_{a} {\partial_{x_a}  y \partial_{x_a} y})^{\frac{3}{2}}}
\left( \left( - \sum_{a}\partial_{x_a} \partial_{x_a} y \right)
\right. \\
& \left(1+\sum_{a} {\partial_{x_a}  y \partial_{x_a} y}\right)
\left.+\sum_{a,b} (\partial_{x_a} y) (\partial_{x_b} y) (\partial_{x_a} \partial_{x_b} y) \right),
\end{split}
\end{equation}
and
\begin{equation}\label{order1eqsp}
 \frac{(d+1) T^{d+1} \sum_a (w_a^2 P_a \ell_0 x_a)}{\left(1-\sum_a (w_a P_a \ell_0)^2\right)^{\frac{d+3}{2}}}
+ \frac{1}{(1+\sum_{a}{\partial_{x_a}  y \partial_{x_a} y})^{\frac{1}{2}}}
\left( \frac{1}{\ell_0 P_1} + \sum_a \frac{\partial_{x_a} y}{\ell_0 P_{(a+1)}}
\right) =0.
\end{equation}
As mentioned before here the sums run from 2 to $n$.
Also in this case besides equating the coefficients of $x_a$s to zero in \eqref{order1eqsp} we also have
set the coefficient of $(R^2 - \sum_a x_a^2)^{\frac{1}{2}}$ to zero to obtain \eqref{genres2}.

In \eqref{genres2} we find a very curious fact about the speed of the class of
solutions that we analyze here. The velocities $w_a P_a\ell_0$ reach a maximum when $R \rightarrow 0$
and the maximum value is given by
\begin{equation}\label{maxspeed:eq}
 w_a^{max} P_a \ell_0 = \frac{1}{\sqrt{(d-n-2)(d+1)+n}}.
\end{equation}
This value is consistent with the  maximum speed for the ring in $3+1$ dimensions
quoted in \cite{Bhardwaj:2008if}. Also note that for the ring ($n=1$) and  large $d$ this maximum
value goes as $\frac{1}{d}$;
this is unlike (although consistent with) the behavior of the ring in asymptotically flat space
where this goes as $\frac{1}{\sqrt{d}}$ (see \cite{Emparan:2007wm, Emparan:2008eg}).
However, this limiting value occurs when $R \to 0$, where our approximation of the surface as
having no thickness breaks down. Nevertheless, for large space time
dimension the behavior of black holes in asymptotically flat spaces and that in
asymptotically AdS spaces are  expected to be similar (see \cite{Caldarelli:2008mv}).
In the light of this fact we may conclude that our fluid approximation is unable to
capture this phenomenon correctly, unless there actually exist a better bound in flat
space (which would go as $\frac{1}{d}$ for large $d$).

In the coordinates that we have used in this section, the first derivative of our solution
$\left( \partial_b y(x_a) = \frac{-x_b}{(R^2 - \sum x_a^2)^{\frac{1}{2}}}\right)$ is
singular near $\sum x_a^2 = R^2$.
In fact (from our analysis so far) it is unclear whether a consistent perturbation theory can be performed in
the $\epsilon$ parameter about the solution \eqref{gensol}. In the later sections we shall move to
better coordinates and exhibit the existence of a well controlled perturbation theory about \eqref{gensol}.
In certain special cases we shall explicitly compute the first correction
 to \eqref{gensol} which occurs at $O(\epsilon^2)$.

\subsection*{Summary\footnote{While this work was being completed we came to know that similar results
were also obtained by Camps, Emparan and Haddad \cite{Emparan:FW}.}}

Here we have constructed a class of fluid configurations to the $d$ dimensional Navier-Stokes equation
in the generalized thin ring limit. To leading order in the parameter $\epsilon$ these
fluid configurations are given by,
\begin{equation*}
 B^{(d-1-n)} \times T^n = B^{(d-1-n)} \times \underbrace{ S^1 \times S^1 \dots  \dots S^1}_{\text{n times}},
\end{equation*}
where $n \leq [\frac{d-1}{2}]$.
These configurations are parameterized by the radius of the ball ($R$) and
the radii of the various $S^1$s ($\ell_0 P_a$). In the generalized thin ring
limit locally these configurations are like filled cylinders with the
topology $B^{(d-1-n)} \times \R^n$. Then we can bend the different
directions in $\R^n$ into $S^1$s in a controlled way with a perturbation expansion in $\epsilon$.
Now the intrinsic fluid parameters (namely the temperature ($T$) and the angular velocities ($\omega_a$))
are related to the parameters of the fluid configuration ($R$ and $\ell_0 P_a$) by the force balance
conditions. The pressure along the radial direction of the ball is balanced by the surface tension.
This condition yields
\begin{equation*}
 T^{d+1} = \left( \frac{(d-n-2)+R}{R} \right)
		\left( 1 - \sum_{a} (\ell_0 P_a w_a)^2\right)^{\left(\frac{d+1}{2}\right)}\\
\end{equation*}
On the other hand the pressure along the radial direction of the $S^1$s is balanced by
the centrifugal force. In order to obtain this force balance we require these
configurations to be rotating (at least) in the planes in which the $S^1$s lie.
For the sake of simplicity we have turned off angular velocity along any other
direction. This force balance determines the angular velocities to be
\begin{equation*}
 w_a^2 = \frac{1}{\left(\ell_0 P_a \right)^2\left(((d-n-2)+R)(d+1)+n\right)}.
\end{equation*}

These fluid configurations are dual to the horizon topology $S^{(d-n)} \times T^n$.
Thus by exploiting the AdS/CFT correspondence we indirectly confirm the existence of such
exotic horizon topologies. It is possible to generate a perturbation expansion (in the
$\epsilon$ parameter) about these solutions as we shall demonstrate in some explicit
examples in the later sections. Also the thermodynamic properties of the fluid configurations
directly map to that of these exotic black objects. This provides us with a opportunity
to study the thermodynamics of these black objects without performing a direct gravity calculation.

\section{Rings}\label{sec:rings}

In this section we shall analyze the topology $B^{d-2} \times S^1$. This topology
is the special case of \eqref{topo:eq} (for $n=1$). We shall study this ring type
solutions in $3+1$ and $4+1$ dimensions in detail. We use regular coordinates to
set up a well controlled perturbation expansion in the parameter $\epsilon$
to compute corrections to the {\it thin ring}. The thin ring solution in
$3+1$ dimensions has been well studied (including the thermodynamic properties)
in \cite{Bhardwaj:2008if}. Here we present the first correction to that
solution demonstrating the fact that it is possible to construct such rings as
a series in the $\epsilon$ parameter. This method can in principle be
generalised to higher dimensions (we present the results in $4+1$
dimensions in \S\ref{sec:rings5d}). All these solutions  are
consistent with the general case discussed in \S \ref{sec:arbdim} to zeroth order in $\epsilon$.

\subsection{Rings in 3+1 dimensions}\label{sec:rings4d}

As pointed out previously in \S\ref{sec:arbdim} the coordinates that we used for our general discussion
(which are those that are used in \cite{Lahiri:2007ae,Bhardwaj:2008if}) are not suitable for the perturbation theory. Therefore,
for the construction of a regular well controlled perturbation theory, we move to some regular coordinates.

Here we use the coordinates $\{t,\rho, \theta, \phi\}$, $t$ being the time coordinate, which are related to those used previously by
\begin{equation}\label{newcoord4d:eq}
\begin{aligned}
  r &= \mathcal{L} + \rho \cos\theta ,&\qquad
  z &= \rho \sin\theta.
\end{aligned}
\end{equation}
The metric is given by
\begin{equation}\label{newmet4d:eq}
 \dr s^2 = - \dr t^2 + \dr\rho^2 + \rho^2 \dr\theta^2
           + ( \mathcal{L} + \rho \cos\theta )^2 \dr\phi^2.
\end{equation}
Here $\mathcal{L}$ is a number which we shall determine in terms of the parameters of our system order by order in
perturbation theory. Physically $\mathcal{L}$ is the distance of the center of the cylinder (our 0th order solution)
from the origin. We simply redefine this center to be our origin in the above metric.

\begin{figure}[htbp]
\begin{center}
\input{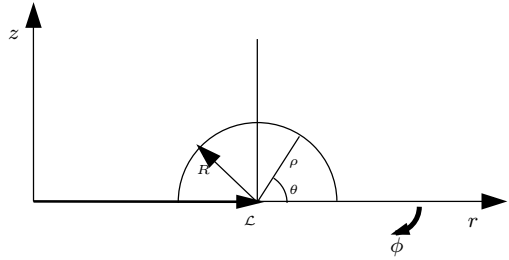}
\end{center}
\caption{Cross section of the 3+1 dimensional ring. The curved arrow labelled $\phi$ indicates a direction that has been suppressed.}
\label{fig:4dring}
\end{figure}

These coordinates are described in fig.\ref{fig:4dring}.

The velocity vector is given by $u^{\mu} = \gamma(1,0,0,\omega)$, where again
$\gamma = \left(1-\omega^2 (\mathcal{L} + \rho \cos(\theta))^2 \right)^{-\frac{1}{2}}$ is the normalization constant.
In this case we consider the fluid surface to be given by
\begin{equation*}
 f \equiv g(\theta) - \rho = 0.
\end{equation*}

Then \eqref{rotsurfbc:eq} reduces to the following differential equation for the function $g(\theta)$
\begin{equation}\label{4dmaineq}
\begin{split}
& \frac{T^5}{\left(1-\omega ^2 (\mathcal{L}+\cos (\theta ) g(\theta
   ))^2\right)^{5/2}}-1  -\frac{1}{(\mathcal{L}+\cos (\theta ) g(\theta )) \left(g(\theta
   )^2+g'(\theta )^2\right)^{3/2}}\left(\sin (\theta ) g'(\theta
   )^3 \right.\\ &\left. +(2 \mathcal{L}+3 \cos (\theta ) g(\theta )) g'(\theta
   )^2+g(\theta )^2 \sin (\theta ) g'(\theta )+g(\theta )
   \left(g(\theta ) (\mathcal{L}+2 \cos (\theta ) g(\theta
   )) \right. \right. \\  & \left. \left.  -(\mathcal{L}+\cos (\theta ) g(\theta )) g''(\theta
   )\right) \right) = 0.
\end{split}
\end{equation}

Now we perform the following scaling:
\begin{equation}\label{scale4d:eq}
 \begin{split}
  \omega & = \epsilon \ w \\
  g(\theta) &= R + \epsilon g_1(\theta) + \epsilon^2 g_2(\theta) + \epsilon^3 g_3(\theta) \\
  \mathcal{L} &= \frac{1}{\epsilon} \ell_0 +  \ell_1 + \epsilon \ell_2 + \epsilon^2 \ell_3
 \end{split}
\end{equation}
where $\epsilon$ is the small parameter with which we wish to perform the perturbation.
Here $g_1, g_2, g_3$ are functions to be determined and $\ell_0, \ell_1, \ell_2, \ell_3$ are
to be expressed in terms of the fluid parameters. Also note that to zeroth order $g(\theta) = R$,
which is the same solution that has been obtained in \cite{Bhardwaj:2008if}. Also the solution in \cite{Bhardwaj:2008if}
(as in the general discussion above) was true up to first order in $\epsilon$. This implies that
the first order correction to $g(\theta)$ (i.e. $g_1(\theta)$) should vanish, as we shall shortly
show.

To first order in $\epsilon$  \eqref{4dmaineq} reduces to
\begin{equation}\label{cond1}
 -1 - \frac{1}{R} + \frac{T^5}{(1- w^2 \ell_0^2)^\frac{5}{2}} = 0.
\end{equation}

To higher order in $\epsilon$ we obtain differential equations for $g_1, g_2, g_3$ etc. These
differential equations are of the general form
\begin{equation}
 g_i(\theta) + g_i''(\theta) = \mathcal{S}_j(\theta).
\end{equation}
where $\mathcal{S}(\theta)$ is the source which is determined at a particular order once the
the complete solution up to one lower order is completely known. Also note that the
homogeneous part of the equation is the same at all orders.

The equation that we obtain at first order is
\begin{equation}
 g_1(\theta) + g_1''(\theta) = -R^2 \left(\frac{5 T^5 \ell_0 \ell_1 w^2}{\left(1-\ell_0^2
   w^2\right)^{7/2}}+\left(\frac{5 T^5 \ell_0 R
   w^2}{\left(1-\ell_0^2
   w^2\right)^{7/2}}-\frac{1}{\ell_0}\right) \cos
   (\theta ) \right)
\end{equation}
Solving the above equation we obtain
\begin{equation}\label{g1exp}
\begin{split}
 g_1(\theta) = &\mathcal{C}_1 \cos(\theta) + \mathcal{C}_2 \sin{\theta} \\
   &+ \frac{R^2 \left(\left(\left(1-\ell_0^2 w^2\right)^{7/2}-5 T^5 \ell_0^2 R w^2\right) (\cos (\theta )
   +2 \theta  \sin (\theta ))-20 T^5 \ell_0^2 \ell_1 w^2\right)}{4 \ell_0 \left(1-\ell_0^2 w^2\right)^{7/2}}
\end{split}
\end{equation}
where the $\mathcal{C}_1$ and $\mathcal{C}_1$ are the integration constants to be determined by the
boundary conditions. We must also remember that the constants $R, \ell_o, \ell_1 $ are also to be determined
in terms of the other fluid parameters, viz.\ $K$ and $w$.

Now from physical considerations we shall demand that the surface should be a closed surface.
This results in the boundary condition
\begin{equation}\label{bdycond}
 g_1'(0) = 0 ; \quad g_1'(\pi) = 0.
\end{equation}
The first condition demands $\mathcal{C}_2 = 0$, while the second condition yields
\begin{equation}\label{cond2}
 \frac{5 T^5 \ell_0^2 R w^2}{\left(1-\ell_0^2 w^2\right)^{7/2}}-1 = 0.
\end{equation}
Here the relations \eqref{cond1} and \eqref{cond2} may be used to express
$R$ and $\ell_0$ in terms of the fluid parameters $T$ and $w$. However for
performing the calculations it is more convenient to do the reverse.
We then have
\begin{equation}\label{4dpar}
 \begin{aligned}
  T^5 &= \left( 1 + \frac{1}{R} \right) \left(\frac{5(1+R)}{6+ 5 R}\right)^\frac{5}{2}, &\qquad
  w &= \frac{1}{\ell_0 \sqrt{6 + 5R }}.
 \end{aligned}
\end{equation}

Plugging back these relations into \eqref{g1exp} we find $g(\theta)$ up to order $\epsilon$,
which is given by
\begin{equation}
 g(\theta) = R + \epsilon \left( \mathcal{C}_1 \cos(\theta) - \frac{\ell_1 R}{\ell_0}\right) + O(\epsilon^2),
\end{equation}
where $\mathcal{C}_1$ and $\ell_1$ are still to be determined.

Now once we start including corrections we should consider a redundancy in description of
the ring which we have to remove by proper gauge fixing.
This pertains to the fact that we haven't defined $\mathcal{L}$ properly yet. Vaguely, it is $r$ coordinate of the center of the ring. This
becomes ill-defined once we take into consideration the corrections to the thin ring, which was a circle in the $\rho$-$\theta$ plane to zeroth order.
This is taken care by the following gauge-fixing condition
\begin{equation}\label{gfix2}
 \int_0^{\pi} g(\theta) \cos(\theta) = 0.
\end{equation}
In words, this condition states that the average $r$ coordinate of the surface in the $r$-$z$ plane is $\mathcal{L}$. This condition implies $\mathcal{C}_1 = 0$.
The constant $\ell_1$ will only be determined at next order in epsilon just as $\ell_0$ was determined
at $\CO(\epsilon)$. Therefore we now proceed to the calculation of the second order corrections to $\epsilon$.

The equation for $g_2(\theta)$ is given by the coefficient of $\epsilon^2$ in \eqref{4dmaineq}. After
plugging in the solution for $g_1(\theta)$ we obtain
\begin{equation}
\begin{split}
 g_2(\theta) + g_2''(\theta) = -\frac{R}{10 \ell_0^2  (R+1)} \left((2-5 R) \ell_1^2 \right.&+ 10 \ell_0 \ell_2 (R+1)
 \\&\left. +R \cos (\theta )   (2 \ell_1 (5 R+12)+R (15 R+22) \cos (\theta ))\right).
\end{split}
\end{equation}

The solution to the above equation is given by
\begin{equation}\label{g2exp}
\begin{split}
 g_2(\theta) = &\mathcal{C}_3 \cos(\theta) + \mathcal{C}_4 \sin{\theta} \\
   &+ \frac{R}{60 \ell_0^2 (R+1)}  \left(R (-3 \ell_1 (5 R+19) \cos (\theta )+R
   (15 R+22) \cos (2 \theta )\right. \\ & \quad \quad \left.-6 \ell_1 (5 R+12) \theta
    \sin (\theta ))-3 \left((4-10 R) \ell_1^2+20
   \ell_0 \ell_2 (R+1)+R^2 (15
   R+22)\right)\right)
\end{split}
\end{equation}
where again $\mathcal{C}_3$ and $\mathcal{C}_4$ are integration constants to be determined.
The boundary conditions (i.e. $g_2'(0)=0=g_2'(\pi)$) again imply $\mathcal{C}_4 = 0$ and the condition
\begin{equation}
 \frac{\ell_1 \pi  R^2 (5 R+12)}{10 \ell_0^2 (R+1)} = 0.
\end{equation}
The above condition implies $\ell_1 = 0$. At this point the $\epsilon$ order solution is completely determined;
and we find that corrections to $g(\theta)$ and $\mathcal{L}$ all vanish. However we go ahead further to
compute the $\CO(\epsilon^2)$ correction completely as it will provide us with the leading order correction.
Using the fact $\ell_1 = 0 $ and $\mathcal{C}_4 = 0$ we find $g(\theta)$ to be
\begin{equation}
\begin{split}
 g(\theta) = R +&  \mathcal{C}_3 \cos(\theta) \epsilon^2 \\ & + \frac{\left(+R
   \left(R^2 (15 R+22) \cos (2 \theta )-3 \left((15 R+22)
   R^2+20 \ell_0 \ell_2 (R+1)\right)\right)\right)
  }{60 \ell_0^2 (R+1)}  \epsilon ^2 + O(\epsilon)^3.
\end{split}
\end{equation}
Now  the condition \eqref{gfix2} again implies $\mathcal{C}_3 =0$.
Again in order to determine $\ell_2$
we have to perform one higher order calculation. We can then determine $\ell_2$
by imposing the boundary conditions on  $g_3(\theta)$. Here we intend to present
only the leading order corrections and therefore do not specify the details of the
third order calculation. However the value of $\ell_2$ that we obtain is given by
\begin{equation}
 \ell_2 = \frac{R^2 \left(225 R^2+380 R+92\right)}{40 \ell_0
   \left(5 R^2+17 R+12\right)}.
\end{equation}
Note that the denominator in the above expression never vanishes for
positive values of $R$.

In summary we can write
\begin{equation}\label{4dringcor}
 \begin{split}
  g(\theta) &= R+\frac{R^3 (2 (5 R+12) (15 R+22) \cos (2 \theta )
-15 (3 R (25 R+64)+124)) \epsilon ^2}{120 \ell_0^2 (R+1) (5 R+12)}+O\left(\epsilon ^3\right)\\
  \mathcal{L} &= \frac{1}{\epsilon} \ell_0 +  \epsilon \frac{R^2 \left(225 R^2+380 R+92\right)}{40 \ell_0
   \left(5 R^2+17 R+12\right)} +O(\epsilon^2)
\end{split}
\end{equation}
As mentioned earlier $R$ and $\ell_0$ are related to the fluid parameters $T$ and $w$ through the
inverse of the relations \eqref{4dpar}. We can carry forward to arbitrary order in $\epsilon$
in a perfectly well controlled fashion.

Finally we would like to mention that it is possible to obtain corrections to the radius $R$
from \eqref{4dringcor}. However, just like the center, the notion of the radius has to be redefined
for the corrected solution. This we may do by defining an average corrected radius,
\begin{equation}\label{Ravg}
 \Ra = \int_{0}^{2\pi} g(\theta) d\theta.
\end{equation}
Using \eqref{4dringcor} we can compute $\Ra$ to be
\begin{equation}
 \Ra = R-\frac{\left( R^3(3R(25R+64)+124) \right) \epsilon^2}
              {8 \left(\ell_0^2 (R+1) (5R+12)\right)}
              +\CO\left(\epsilon ^3\right).
\end{equation}

\begin{figure}[htbp]
\begin{center}
  \begin{center}
  \input{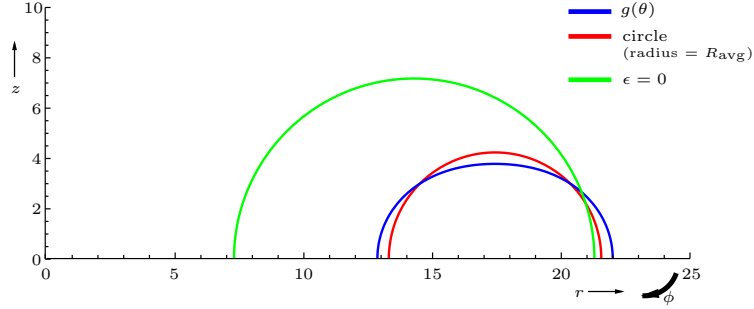}
  \end{center}
  \caption{A plot of the corrected solution \eqref{4dringcor} for the 3+1 dimensional ring with $\frac{\ell_0}{\epsilon}=14$ and $R=7$.}\label{fig:4dringplot}
\end{center}
\end{figure}

We present a plot of this corrected solution in fig.\ref{fig:4dringplot}.

The thermodynamic properties of the solution can be computed from \eqref{rotthermscaled:eq} as follows
\begin{multline}\label{therm4dring:eq}
  T\ln\gpf = 2\pi
\left [
  \int_0^\pi\!\!\dr\theta \int_0^{g(\theta)}\!\!\!\!\dr\rho\,
     \rho(\mathcal{L}+\rho\cos\theta)(K\gamma^5-1)
\right. \\  \left.
  -\int_0^\pi\!\!\dr\theta
    \sqrt{g(\theta)^2+g'(\theta)^2}(\mathcal{L}+g(\theta)\cos\theta)
\right ]
\end{multline}

We find
\begin{equation}\label{thermexp4dring:eq}
\begin{split}
  T\ln\gpf &=
   -\frac{2 \pi^2 \ell_0 R}{\epsilon}
   +\frac{2 \pi^2 R^3 (15 R+22)}{40 \ell_0 (R+1)} \,\epsilon
   +\CO\left(\epsilon^2\right), \\
\end{split}
\end{equation}
The other thermodynamic properties can be found by differentiating this \eqref{gpftherm:eq}:
\begin{equation}\label{thermdif4dring:eq}
\begin{split}
  E &=
   \frac{2 \pi ^2 \ell_0 R (5 R+7)}{\epsilon }
   -\frac{2 \pi^2R^3(R(5R(105R+353)+2018)+792)}{8(\ell_0(R+1)(5 R+12))} \,\epsilon
   +\CO\left(\epsilon^2\right) ,\\
  S &=
   \frac{2 \pi^2 \sqrt{5(5R+6)} \ell_0 R^{6/5} (R+1)^{3/10}}{\epsilon}\\
   & \quad \quad \quad -\frac{2 \pi^2 R^{16/5} \sqrt{5 R+6} (R(5R(105R+353)+2018)+792)}
   {8 \sqrt{5} \ell_0 (R+1)^{17/10} (5R+12)} \,\epsilon
   + \CO\left(\epsilon^2\right) ,\\
  L &=
   \frac{2 \pi^2 \ell_0^2 R \sqrt{5 R+6}}{\epsilon^2}
   +\frac{2 \pi^2 R^3 \sqrt{5R+6} (15R+22)}{40(R+1)}
   +\CO\left(\epsilon^1\right) .
\end{split}
\end{equation}
Note that to leading order these expressions for energy, entropy and angular momentum
match with that presented in \cite{Bhardwaj:2008if} after performing the
necessary variable transformations.

\begin{figure}[htbp]
  \begin{center}
  \input{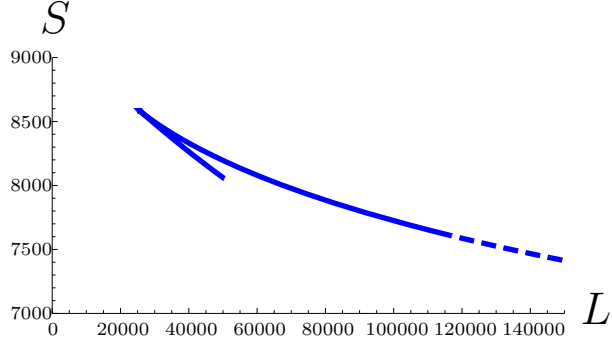}
  \end{center}
  \caption{Plot of entropy, $S$, vs.\ angular momentum, $L$, at fixed energy, $E=10,000$, for the ring in 3+1 dimensions. The dotted portion of the
curve represents the region where $R < 1$ and hence lie outside the surface tension approximation.}\label{fig:ring4dLvsS}
\end{figure}

We present a plot of the thermodynamic properties of this ring in fig.\ref{fig:ring4dLvsS}

\subsection{Rings in 4+1 dimensions}\label{sec:rings5d}

We now analyze the ring in one higher dimension i.e. $4+1$ dimension. The construction is
exactly parallel to that in $3+1$ dimensions and therefore we skip most of the details and specify
only the result.

The coordinates that we use here are $\{t,\rho,\theta,\phi_1,\phi_2\}$, which are related to the old coordinates by
\begin{equation}\label{newcoord5dring:eq}
\begin{aligned}
  r_1 &= \mathcal{L} + \rho \cos\theta ,&\qquad
  r_2 &= \rho \sin\theta,
\end{aligned}
\end{equation}
with the metric
\begin{equation}\label{newmet5dring:eq}
 \dr s^2 = -\dr t^2 + \dr \rho^2 + \rho^2 \dr\theta^2
   + (\mathcal{L} + \rho \cos\theta)^2 \dr\phi_1^2 + (\rho \sin\theta)^2 \dr\phi_2^2.
\end{equation}
These coordinates are described in fig.\ref{fig:5dring}.

\begin{figure}[htbp]
\begin{center}
\input{5dring.TpX}
\end{center}
\caption{Cross section of the 4+1 dimensional ring. The curved arrows labelled $\phi$ indicate a direction that has been suppressed.}
\label{fig:5dring}
\end{figure}

The velocity is given by $u^{\mu}= \gamma(1,0,0,\omega,0)$ where the normalization is given by
$\gamma = (1-\omega^2 (\mathcal{L} + \rho \cos(\theta))^2)^{-\frac{1}{2}}$. We take the fluid surface to be
$f \equiv g(\theta) - \rho = 0$. Then the equation \eqref{rotsurfbc:eq} reduces to
\begin{equation}
\begin{split}
 & \frac{T^6}{\left(1-\omega ^2 (\mathcal{L}+\cos (\theta ) g(\theta
   ))^2\right)^3}-1  -\frac{\left(g(\theta )+g''(\theta
   )\right) g'(\theta )^2}{g(\theta )^4
   \left(\frac{g'(\theta )^2}{g(\theta
   )^2}+1\right)^{3/2}} \\ & +\frac{1}{\left(\mathcal{L}+\cos
   (\theta ) g(\theta )\right)\left(g(\theta )^4
   \left(\frac{g'(\theta )^2}{g(\theta
   )^2}+1\right)^{3/2}\right)}\left(g(\theta
   )^2+g'(\theta )^2\right) \left(g(\theta ) (2 \mathcal{L}+3 \cos
   (\theta ) g(\theta ))- \right. \\ &-\left. \csc (\theta ) (\mathcal{L} \cos (\theta
   )+\cos (2 \theta ) g(\theta )) g'(\theta )-(\mathcal{L}+\cos
   (\theta ) g(\theta )) g''(\theta )\right) = 0,
\end{split}
\end{equation}

Now we plug in the scaling with $\epsilon$ as in the previously discussed cases:
\begin{equation}\label{scale5dr:eq}
 \begin{split}
  \omega & = \epsilon \ w \\
  g(\theta) &= R + \epsilon g_1(\theta) + \epsilon^2 g_2(\theta) + \dots \\
  \mathcal{L} &= \frac{1}{\epsilon} \ell_0 +  \ell_1 + \epsilon \ell_2 + \dots
 \end{split}
\end{equation}
Then, performing an analysis exactly the same as the one performed in \S\ref{sec:rings4d},
we obtain
\begin{equation}\label{5dringcor}
 \begin{split}
  g(\theta) &= R+\frac{R^3  \left(5 \left(81 R^2+432 R+572\right)
   \cos (2 \theta )-3 \left(243 R^2+1040
   R+1108\right)\right)}{240 \ell_0^2 \left(9 R^2+44
   R+52\right)} \epsilon ^2 +\CO \left(\epsilon ^3\right)\\
  \mathcal{L} &= \frac{1}{\epsilon} \ell_0 + \frac{R^2 \left(81 R^2+240 R+116\right)}
	{30 \ell_0 \left(9 R^2+44 R+52\right)}\epsilon
        +\CO \left(\epsilon ^2\right),
 \end{split}
\end{equation}
with $R$ and $\ell_0$ being expressed in terms of the fluid parameters $T$
and $w$ (implicitly) by the following relations
\begin{equation}
\begin{aligned}
  T^6 &= \frac{216 (R+2)^4}{R (6 R+13)^3}\,, &\qquad
  w &= \frac{1}{\ell_0 \sqrt{6 R+13}}\,.
\end{aligned}
\end{equation}
Even in this case we note that all the $\CO(\epsilon)$ corrections vanish.

Finally the average radius of the curve in \eqref{5dringcor}
($\Ra$ as defined in \eqref{Ravg}) is given in this case by
\begin{equation}
\Ra = R-\frac{\left(R^3 (243 R+554)\right) \epsilon ^2}{80
   \left(\ell_0^2 (9 R+26)\right)}+ \CO\left(\epsilon ^3\right).
\end{equation}

We present a plot of this corrected solution in fig.\ref{fig:5dringplot}.

\begin{figure}[htbp]
\begin{center}
  \input{5dring_plot.TpX}
  \caption{A plot of the corrected solution \eqref{5dringcor} for the 4+1 dimensional ring with $\frac{\ell_0}{\epsilon}=14$ and $R=7$.}\label{fig:5dringplot}
\end{center}
\end{figure}

The thermodynamic properties of the solution can be computed from \eqref{rotthermscaled:eq} as follows
\begin{multline}\label{therm5dring:eq}
  T\ln\gpf = (2\pi)^2
\left[
  \int_0^\pi\!\!\dr\theta \int_0^{g(\theta)}\!\!\dr\rho\,
     \rho^2\sin\theta(\mathcal{L}+\rho\cos\theta)(K\gamma^6-1)
\right.\\\left.
  -\int_0^\pi\!\!\dr\theta
    \sqrt{g(\theta)^2+g'(\theta)^2}(\mathcal{L}+g(\theta)\cos\theta)g(\theta)\sin\theta
\right]
\end{multline}
We find
\begin{equation}\label{thermexp5dring:eq}
\begin{split}
  T\ln\gpf =& \frac{4 \pi ^2 \ell_0 R^2 }{9 \epsilon }\left(-6
   R+\sqrt[3]{R+2} \sqrt{36 R+78}
   \sqrt[6]{R}-18\right) \\
   &+\frac{2 \pi
   ^2 R^4 }{135 \ell_0 (R+2)^{5/3} (9
   R+26)}\left(-3216 \sqrt{36 R+78} R^{7/6}-243
   \sqrt{36 R+78} R^{19/6} \right. \\
   &\left. -1524 \sqrt{36 R+78}
   R^{13/6}+1458 (R+2)^{2/3} R^3+9630 (R+2)^{2/3}
   R^2\right. \\
   & \left. +21888 (R+2)^{2/3} R-2288 \sqrt{36 R+78}
   \sqrt[6]{R}+17160 (R+2)^{2/3}\right) \epsilon
   \\ &
   +\CO(\epsilon ^2)
   , \\
\end{split}
\end{equation}

\begin{figure}[htbp]
  \begin{center}
  \input{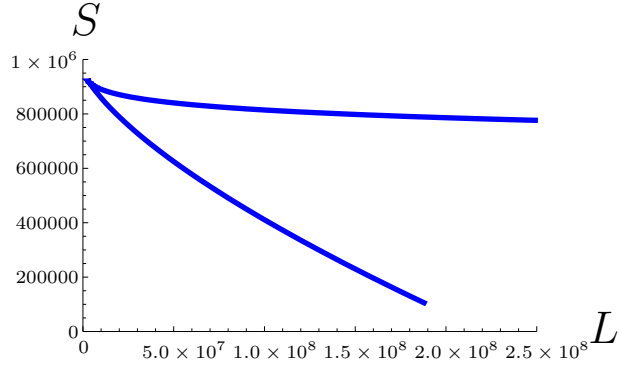}
  \end{center}
  \caption{Plot of entropy, $S$, vs.\ angular momentum, $L$, at fixed energy, $E=1,000,000$, for the ring in 4+1 dimensions.}\label{fig:ring5dLvsS}
\end{figure}

For expressions of the energy, angular momentum and entropy refer to Appendix \ref{sec:largeexp}.
We present a plot of the thermodynamic properties of this ring in fig.\ref{fig:ring5dLvsS}

\section{`Torus' in 4+1 dimension} \label{sec:torus}

Here we analyze the solution with the topology $B^2 \times S^1 \times S^1$ which for the lack of terminology we
refer to as the `torus'. Although the perturbation
theory for the torus is almost exactly parallel to that for the ring, however there are certain
differences as far as imposing the boundary condition is concerned. We consider
spatial part of the 4+1 dimensional space to consist of 2 independent planes.
We turn on two independent angular velocities $(\omega_1, \omega_2)$ along a direction orthogonal to these planes.
The two $S^1$s of the topology $B^2 \times S^1 \times S^1$ lie on these two planes.
Initially we put polar coordinates $r_1,\phi_1$ and $r_2,\phi_2$  on these two planes. Further
in the $(r_1,r_2)$ plane we shift to coordinates $\rho, \theta$ after a shift  in the origin by
the vector $\overrightarrow{\mathcal{L}} = (\mathcal{L}\cos(\chi), \mathcal{L}\sin(\chi))$ (expressed in the $(r_1,r_2)$ coordinates).
The various coordinates have been represented in fig.\ref{fig:5dtorus}.
Thus finally we work with the coordinates $\{t, \rho, \theta, \phi_1, \phi_2\}$, which are related to the old coordinates by
\begin{equation}\label{newcoord5dtorus:eq}
\begin{aligned}
  r_1 &= \mathcal{L} \cos\chi+ \rho \cos\theta, &\qquad
  r_2 &= \mathcal{L} \sin\chi+ \rho \sin\theta,
\end{aligned}
\end{equation}
with the metric
\begin{equation}\label{newmet5dtorus:eq}
 \dr s^2 = -\dr t^2 + \dr\rho^2+ \rho^2 \dr\theta^2
                    + (\mathcal{L} \cos\chi+ \rho \cos\theta)^2 \dr\phi_1^2
					+ (\mathcal{L} \sin\chi+ \rho \sin\theta)^2 \dr\phi_2^2
\end{equation}

\begin{figure}[htbp]
\begin{center}
\input{5dtorus.TpX}
\end{center}
\caption{Cross section of the 4+1 dimensional torus. The curved arrows labelled $\phi$ indicate a direction that has been suppressed.}
\label{fig:5dtorus}
\end{figure}

Here the velocity is given by $u^{\mu} = \gamma(1, 0, 0, \omega_1,\omega_2)$ where again the
normalization is given by
$\gamma = (1- (\mathcal{L} \cos\chi+ \rho \cos\theta)^2 \omega_1^2 -(\mathcal{L} \sin\chi+ \rho \sin\theta)^2 \omega_2^2)^{-\frac{1}{2}}$.
We consider the fluid surface to be given by $f \equiv g(\theta) - \rho =0$.

Now the differential equation satisfied by $g(\theta)$ (which is \eqref{rotsurfbc:eq} for the present case) is given by
\begin{equation}
\begin{split}
&  \frac{T^6}{\left(1 - \omega_1^2 (\mathcal{L}\cos\chi+g(\theta)\cos\theta)^2
          - \omega_2^2 (\mathcal{L}\sin\chi+ g(\theta)\sin\theta)^2\right)^3}
   -1
   -\frac{\left(g(\theta)+g''(\theta)\right) g'(\theta)^2}
   {g(\theta)^4 \left(\frac{g'(\theta)^2}{g(\theta)^2}+1\right)^{3/2}}\\
&+ \frac{\left(g(\theta )^2+g'(\theta )^2\right)}{g(\theta )^2
   \left(\frac{g'(\theta )^2}{g(\theta
   )^2}+1\right)^{3/2}}
   \left(\frac{\cos (\theta ) g'(\theta )-g(\theta ) \sin
   (\theta )}{\sin (\theta ) g(\theta )^2+\mathcal{L} \sin (\chi )
   g(\theta )}-\frac{\cos (\theta ) g(\theta )+\sin
   (\theta ) g'(\theta )}{\cos (\theta ) g(\theta )^2+\mathcal{L}
   \cos (\chi ) g(\theta )} \right.  \\ & \left. +\frac{g''(\theta )-g(\theta
   )}{g(\theta )^2}\right) = 0,
 \end{split}
\end{equation}

Now we again consider the following scaling:
\begin{equation}\label{scale5dt:eq}
 \begin{split}
  \omega_1 &= \epsilon \ w_1 \\
  \omega_1 & = \epsilon \ w_2 \\
  g(\theta) &= R + \epsilon\  g_1(\theta) + \epsilon^2 \ g_2(\theta)  + \dots\\
  \mathcal{L} &= \frac{1}{\epsilon} \ell_0 +  \ell_1 + \epsilon \ \ell_2 + \dots \\
  \chi &= \chi_0 + \epsilon \ \chi_1 + \epsilon^2 \ \chi_2 + \dots
 \end{split}
\end{equation}
We shall determine the unknown functions in a similar way as we did for the ring.
However there is a crucial difference between the two. Firstly here we have one more
parameter (since we have two angular velocities instead of one). Secondly the
boundary condition that we have to impose on $g(\theta)$ is different from the
previous case because here we are dealing with a different closed surface.
Although physically it is the same criterion -- the fact that we should have a closed
surface, the mathematical formulation of the statement is different as we shall now describe.

Instead of the boundary condition \eqref{bdycond} we should use the condition
\begin{equation}
 \label{bdycondtor}
 g(0) = g(2 \pi); \quad g'(0)=g'(2\pi).
\end{equation}
which is the statement that we should have a closed curve in the $\rho$-$\theta$ plane and that
the curve must close in a regular fashion such that the derivatives on either side of the point
of closing (which we take to be $\theta =0$) are equal. As the differential equation is second order and periodic, this ensures
that all higher derivatives are continuous at $\theta =0$ and the solution is fully periodic.
Besides the boundary conditions we will also have to fix the ambiguity regarding the center
of the torus just as we did for the ring. However unlike the ring the center here does not
lie on the $r_2$ axis. Therefore in order to fix the center we will have to use two conditions namely
\begin{equation}
\begin{aligned}
 \int_0^{2\pi} g(\theta) \cos(\theta) &= 0 ,&
 \int_0^{2\pi} g(\theta) \sin(\theta) &= 0 .
\end{aligned}
\end{equation}
In words, these conditions states that the average $r_1$ coordinate of the surface in the $r_1$-$r_2$ plane is $\mathcal{L}\cos\chi$ and the average $r_2$ coordinate  is $\mathcal{L}\sin\chi$. Then proceeding in the same way as the ring (after including the above modifications)
we find the following result:
\begin{equation}\label{5dtoruscor}
\begin{split}
g(\theta) &= R-\frac{1}{72 \left(\ell_0^2 (R+1)\right)}\left(R^3 \left(9 \left(27 R^2+74 R+51\right)+36 (R+1) \cos
   (2 \theta -2 \chi_0) \right. \right. \\ &\left. \left. +4 (9 R+17) \cos (2 (\theta
   +\chi_0))\right) \csc ^2(2 \chi_0)\right)
   \epsilon ^2+O\left(\epsilon
   ^3\right)\\
\mathcal{L} &= \frac{\ell_0}{\epsilon}+  \frac{R^2 \left(81 R^2+150 R+8 \cos (4 \chi_0)+57\right) \csc ^2(2 \chi_0)}{48
   \ell_0 (R+1)} \epsilon+ O(\epsilon^2) \\
\chi &= \chi_0 -\frac{R^2 \cot (2 \chi_0)}{3 \ell_0^2 (R+1)} \epsilon^2 .
\end{split}
\end{equation}
Note that just as for the ring all the order $\epsilon$ corrections vanish.
Here $R$, $\ell_0$ and $\chi_0$  are again given implicitly in terms of the fluid parameters
$T$, $w_1$ and $w_2$ by the following relations
\begin{equation}
\begin{aligned}
 T^6 &= \frac{27 (R+1)^4}{R (3 R+4)^3} \,, &
 w_1 &= \frac{1}{\ell_0 \cos(\chi_0)}\sqrt{\frac{1}{8 + 6 R}} \,,&
 w_2 &= \frac{1}{\ell_0 \sin(\chi_0)}\sqrt{\frac{1}{8 + 6 R}} \,.
\end{aligned}
\end{equation}
Note that this entirely matches the general results \eqref{genres1} and \eqref{genres2} for $d=5$
and $n=2$.

Again in this case, the average radius of the closed curve \eqref{5dtoruscor}
(with $\Ra$ as defined in \eqref{Ravg}) is given by
\begin{equation}
\Ra = R-\frac{\epsilon ^2 \left(R^3 \left(27 R^2+74 R+51\right)
   \csc ^2(2 \chi_0)\right)}{8 \left(\ell_0^2
   (R+1)\right)}+ \CO\left(\epsilon ^3\right).
\end{equation}

\begin{figure}[tbp]
\begin{center}
  \input{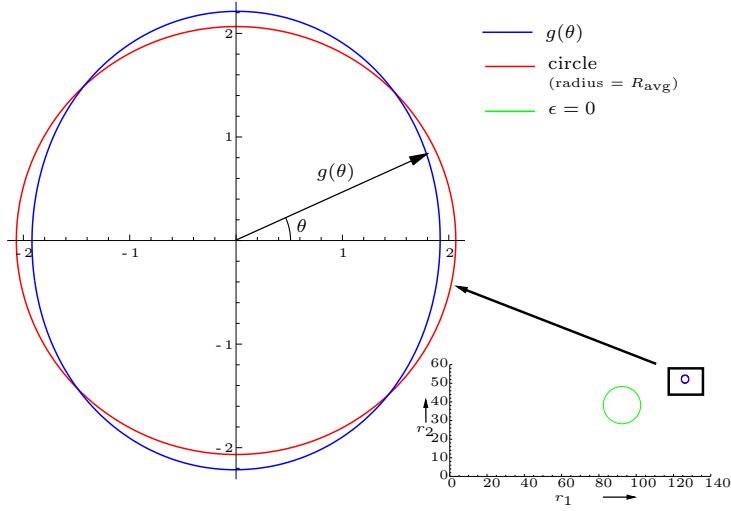}
  \caption{A plot of the corrected solution \eqref{5dtoruscor} for the 4+1 dimensional torus with $\frac{\ell_0}{\epsilon}=100$, $\chi_0=\frac{\pi}{8}$ and $R=10$.}\label{fig:5dtorusplot}
\end{center}
\end{figure}

We present a plot of this corrected solution in fig.\ref{fig:5dtorusplot}.

We expect our construction of the torus solution to break down when
$\ell_0 \sim R$. Which is reflected in the fact that $\frac{g_2(\theta)}{R}$, $\frac{\ell_2}{\ell_o}$ and $\frac{\chi_2}{\chi_0}$
 are all proportional to $\left(\frac{R}{\ell_0}\right)^2$.

The thermodynamic properties of the solution can be computed from \eqref{rotthermscaled:eq} as follows
\begin{multline}\label{therm5dtorus:eq}
  T\ln\gpf = (2\pi)^2
\left[
  \int_0^{2\pi}\!\!\dr\theta \int_0^{g(\theta)}\!\!\!\!\dr\rho\,
     \rho(\mathcal{L}\cos\chi+\rho\cos\theta)(\mathcal{L}\sin\chi+\rho\sin\theta)(K\gamma^6-1)
\right.\\\left.
  -\int_0^{2\pi}\!\!\dr\theta
    \sqrt{g(\theta)^2+g'(\theta)^2}
    (\mathcal{L}\cos\chi+g(\theta)\cos\theta)(\mathcal{L}\sin\chi+g(\theta)\sin\theta)
\right]
\end{multline}
We find
\begin{equation}\label{thermexp5dtorus:eq}
\begin{split}
  T\ln\gpf &=
   \frac{2 \pi ^3 \ell_0^2 R  \sin (2 \chi_0)}
   {3 \epsilon^2}
   \left(-3 R+\sqrt[3]{R+1} \sqrt{9 R+12} \sqrt[6]{R}-6\right)
   \\ &
   +\frac{\pi ^3 R^3  \csc (2 \chi_0)}
   {36 (R+1)^{5/3}}
   (3 (R+1)^{2/3} (R (3 R (27 R+98)+401)+208)\\ &-\sqrt[6]{R} (3 R+4) \sqrt{9 R+12} (R (27 R+62)+39))
   +\CO\left(\epsilon ^1\right)
   , \\
\end{split}
\end{equation}

\begin{figure}[tbp]
  \begin{center}
  \input{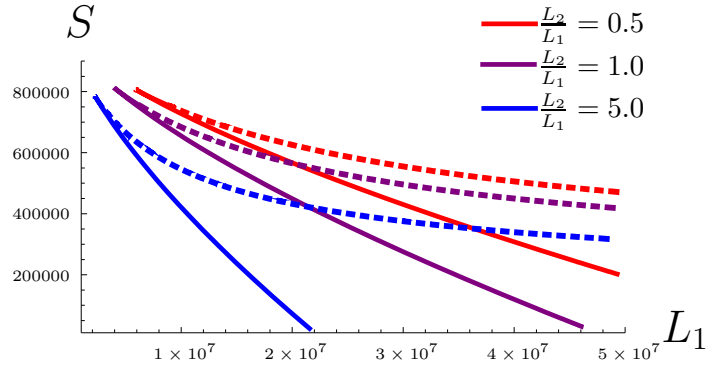}
  \end{center}
  \caption{Plot of entropy, $S$, vs.\ angular momentum, $L_1$, at fixed energy, $E=1,000,000$, and different values of $\frac{L_2}{L_1}$ for the torus in 4+1 dimensions.
Here again the dotted portion of the curve represents the region $R < 1$ which is outside the
validity of the surface tension approximation.}\label{fig:torus5dLvsS}
\end{figure}

Again the expressions for the energy, angular momentum and entropy are given in Appendix \ref{sec:largeexp}.
We present a plot of the thermodynamic properties of this torus in fig.\ref{fig:torus5dLvsS}

\section{Discussion}

In this paper we  have explored an interesting prediction of the AdS/CFT correspondence which connects the dynamics of classical gravity to that of fluid dynamics in appropriate limits. This duality has been exploited by several authors to make interesting gravitational predictions about the properties of the dual fluid (see \cite{Son:2007vk} for a review). In \cite{Bhattacharyya:2008jc}, it was shown that for certain special cases it is possible to reduce Einstein's equations on the bulk side to the Navier-Stokes equations governing the boundary fluid dynamics. In this process they have established a one-to-one correspondence solutions of fluid mechanics and gravitational solutions.

Here we have used this correspondence in a somewhat reverse fashion to prove the existence of a class of very exotic horizon topologies in higher dimensional AdS-like spaces using solutions of the equations of fluid mechanics. However, perhaps due the lack of the corresponding gravitational solutions, we do not have such a precise map between the horizon and the boundary. Nevertheless it is an interesting question worth pursuing in future.

Also we would like to emphasize one interesting observation regarding the velocities of rotation of the solutions that we have studied. In \eqref{maxspeed:eq} we pointed out that maximum velocity of rotation along any of the planes has a upper bound independent of the energy (although the $R \rightarrow 0$ limit in which this bound was obtained lies outside the validity of the surface tension approximation). For the case of the ring (for which $n$=1) this upper bound is given by $\left( (d-3)(d+1)+1 \right)^{-\frac{1}{2}}$ (which for $d=4$ yields $\frac{1}{\sqrt{6}}$  consistent with that obtained in \cite{Bhardwaj:2008if}). Note that for large $d$ this upper bound behaves as $\frac{1}{d}$. We may compare this with the corresponding result in flat space presented in \cite{Emparan:2007wm} where the bound is $(d-1)^{-\frac{1}{2}} \sim \frac{1}{\sqrt{d}}$ for large $d$.

In \cite{Caldarelli:2008mv} the authors had proposed that the black objects in SSAdS spaces with large number of spatial dimensions may qualitatively behave the same way as the corresponding black objects in flat space. This is could possibly be understood as follows: The SSAdS space has one special direction, namely the radial direction, in which Hawking radiation from the plasmaball bounces back. In the other directions, the Hawking radiation escapes to infinity, much like in flat space. As a result, plasmaballs show behaviour that is somehow in between flat space black holes and large AdS black holes. For example, they have a negative heat capacity and eventually evaporate completely. As we increase the number of dimensions, the flat directions outnumber the AdS-like radial direction more and more. Hence, one might expect  the black object living in SSAdS spaces to behave more like their flat space counterparts as one increases the number of dimensions. For example, at fixed energy, plasmarings have an upper bound on the angular momentum in five bulk dimensions, like in AdS \cite{Caldarelli:2008pz}, but we see no such bound in higher dimensions, like in flat space.

With this in view, it might be that there existed a bound for the black objects in SSAdS spaces that went as $\frac{1}{\sqrt{d}}$ for large $d$ and our fluid analysis gives us an incorrect value of this bound (as our approximations breakdown in the limit in which the bound is obtained). Otherwise it might also be that the
qualitative similarity of the fact that a universal upper bound actually exists in both cases is all that
we should expect, there being a difference in the exact value of this bound.

It might also be interesting to study the time dependent fluctuations on these solutions to understand their stability properties. Like many fluid configurations these might suffer from Plateau-Rayleigh instability that would indicate Gregory-Laflamme type instability on the gravity side \cite{Gregory:1993vy, Gregory:1994bj} (see \cite{Caldarelli:2008mv,Maeda:2008kj,Caldarelli:2008ze,Cardoso:2009bv} for recent work connecting the
Plateau-Rayleigh and Gregory-Laflamme instabilities). Since Plateau-Rayleigh type instability is very
commonly found in narrow strips of moving fluid (like the one that we consider here) it may be true that
with our methods we may never be able to capture stable configurations.

In five dimensional gravity there are a number of interesting solutions with disconnected horizons, such as black Saturns \cite{Elvang:2007rd}, with a black hole surrounded by a black ring, and di-rings \cite{Gauntlett:2004wh, Gauntlett:2004qy, Iguchi:2007is, Evslin:2007fv}. These solutions play a significant role in the phase diagram \cite{Elvang:2007hg}. We note that the construction of such solutions in fluid mechanics is trivial -- disconnected lumps of plasma do not affect each other at all. This may sound surprising from the gravitational perspective, as one would expect black holes to attract each other. However, in the AdS soliton background the spectrum of the graviton has a gap $\Lambda_\mathrm{gap} \sim \xi^{-1}$. Therefore one would expect the gravitational attraction to die off exponentially rather than the usual power law, $\e^{-\Lambda_\mathrm{gap}r} \sim \e^{-r/\xi}$. As the surface tension approximations is only valid when the separation  of these lumps of plasma is much larger than the surface thickness, $r \gg \xi$, the gravitational attraction would be negligible.

One issue concerning Saturns is thermal equilibrium between the components \cite{Elvang:2007hg}. This would require the temperature and angular velocity of the components to be equal (see also \cite{Evslin:2008py}). It was pointed out in \cite{Evslin:2008py} that such solutions do not exist in 2+1 dimensions in this framework. However, demanding thermal equilibrium between the disconnected components is only reasonable when the time scale of energy/angular momentum exchange between the components is much shorter than the rate of Hawking leakage to infinity. This limit may not be compatible with the demand that the separation of the components is much larger than the surface thickness.

\section*{Acknowledgements}

We would like to thank Shiraz~Minwalla for his constant guidance throughout
the project. We would also like to thank Roberto~Emparan for useful
discussions. SL would like to thank TIFR for their hospitality when
this work was started. JB would like to thank everyone
in the TIFR theory students room, especially Sayantani~Bhattacharyya
and R.~Loganayagam.


\section*{Appendices}
\appendix

\section{Extrinsic curvature}\label{sec:extrinsic}

Suppose we have a timelike surface with unit normal vector $n$ pointing toward us (spacelike surfaces will require some sign differences). The induced metric on the surface is
\begin{equation}\label{indmet:eq}
  h_{\mu\nu} = g_{\mu\nu} - n_\mu n_\nu.
\end{equation}
The extrinsic curvature is given by \cite{Wald-GeneRela:84}
\begin{equation}\label{extrdef:eq}
  \Theta_{\mu\nu} = \half \CL_n h_{\mu\nu} = \nabla_\mu n_\nu.
\end{equation}
We have to be a little careful with the last expression. It agrees with the first expression when projected tangent to the surface. The first expression has vanishing components normal to the surface. The normal components of the second expression depend on how we extend $n$ off the surface.

The conventional choice for extending $n$ is as follows: at each point on the surface, construct the geodesic that passes through that point tangent to $n$ and parallel transport $n$ along it. In other words
\begin{equation}\label{geodesic:eq}
  n^\mu \nabla_\mu n^\nu = 0.
\end{equation}
This ensures that the second expression in \eqref{extrdef:eq} has vanishing components normal to the surface. The other normal component, $n^\nu \nabla_\mu n_\nu$, vanishes due to the normalisation of $n$.

For the surfaces given by $f(x)=0$, considered in \S\ref{sec:surface}, the unit normal on the surface is given by
\begin{equation}\label{normonsurf:eq}
  n_\mu = -\frac{\p_\mu f}{\sqrt{\p f \cdt \p f}}.
\end{equation}
However, if we used this vector away from the surface, it would not satisfy \eqref{geodesic:eq}. We could still use either expression in \eqref{extrdef:eq} with this vector --- we would just have to project the second one tangent to the surface. Alternatively, we can use
\begin{equation}\label{normoffsurf:eq}
  n_\mu = -\frac{\p_\mu f}{(\p f \cdt \p f)^{1/2}}
   +\brk{ \frac{\p^\nu\! f\, \nabla_\nu \p_\mu f}{(\p f \cdt \p f)^{3/2}}
      - \frac{\p_\mu f\, \p^\lambda\! f\, \p^\nu\! f\, \nabla_\lambda \p_\nu f}{(\p f \cdt \p f)^{5/2}} } f
   + \CO(f^2).
\end{equation}
The $\CO(f^2)$ terms don't contribute to \eqref{extrdef:eq} or \eqref{geodesic:eq} on the surface. The contribution of the $\CO(f)$ terms on the surface to \eqref{extrdef:eq} are normal to the surface and ensure that $n$ satisfies \eqref{geodesic:eq}.

Either way, on the surface, we get
\begin{equation}\label{extrsurf:eq}
  \Theta_{\mu\nu} = -\frac{\nabla_\mu \p_\nu f}{(\p f \cdt \p f)^{1/2}}
    + \frac{\p_\mu f\, \p^\lambda\! f\, \nabla_\lambda \p_\nu f + \p_\nu f\, \p^\lambda\! f\, \nabla_\lambda \p_\mu f}
           {(\p f \cdt \p f)^{3/2}}
    - \frac{\p_\mu f\, \p_\nu f\, \p^\lambda\! f\, \p^\sigma\! f\, \nabla_\lambda \p_\sigma f}
           {(\p f \cdt \p f)^{5/2}}.
\end{equation}
As this is perpendicular to $n$, it doesn't matter if we contract its indices with the full metric $g_{\mu\nu}$ or the induced metric $h_{\mu\nu}$. We get
\begin{equation}\label{trextrsurf:eq}
  \Theta = \Theta_\mu^\mu = -\frac{\square f}{(\p f \cdt \p f)^{1/2}}
    + \frac{\p^\mu\! f\, \p^\nu\! f\, \nabla_\mu \p_\nu f}
           {(\p f \cdt \p f)^{3/2}}.
\end{equation}
%

\section{Energy, angular momentum, and entropy of the
ring and torus in 4+1 dimensions}\label{sec:largeexp}

The energy, $E$, angular momentum, $L$, and entropy $S$ for the $4+1$ dimensional
ring are obtained from \eqref{thermexp5dring:eq} by differentiation \eqref{gpftherm:eq} and are given by
\begin{equation*}
\begin{split}
  E =& \frac{4 \pi ^2 \ell_0 R^2 }{3 \sqrt{6 R+13} (9 R+26)
   \epsilon }\left(-2379 \sqrt{6}
   \sqrt[3]{R+2} R^{7/6}-180 \sqrt{6}
   \sqrt[3]{R+2} R^{19/6}\right. \\& \left.-1128 \sqrt{6}
   \sqrt[3]{R+2} R^{13/6}+180 \sqrt{6 R+13}
   R^3+1188 \sqrt{6 R+13} R^2+2732 \sqrt{6 R+13}
   R \right. \\ & \left. -1690 \sqrt{6} \sqrt[3]{R+2} \sqrt[6]{R}+2184
   \sqrt{6 R+13}\right)\\ &+\frac{2 \pi ^2 R^4 }{135 \ell_0
   (R+2)^{5/3} \sqrt{6 R+13} (9
   R+26)^3}\left(-2598156
   (R+2)^{2/3} \sqrt{6 R+13} R^6\right. \\ & \left. -34169688
   (R+2)^{2/3} \sqrt{6 R+13} R^5-186252048
   (R+2)^{2/3} \sqrt{6 R+13} R^4\right. \\ & \left. -539837568
   (R+2)^{2/3} \sqrt{6 R+13} R^3-879420672
   (R+2)^{2/3} \sqrt{6 R+13} R^2\right. \\ & \left. -765123840
   (R+2)^{2/3} \sqrt{6 R+13} R \right. \\ & \left. +\sqrt{6} (6 R+13)
   (3 R (R (3 R (3 R (243 R (66
   R+853)+1105784)+9364772) \right. \\ & \left. +44322304)+37098880)+38
   667200) \sqrt[6]{R}  -278403840 (R+2)^{2/3}
   \sqrt{6 R+13}\right) \epsilon
   \\ &
   +\CO(\epsilon ^2)
   ,
\end{split}
\end{equation*}

\begin{equation*}
\begin{split}
  S =& \frac{4 \pi ^2 \ell_0 R^{13/6} }{9 \sqrt[3]{R+2} (9 R+26)
   \epsilon }\left(468
   (R+2)^{2/3} \sqrt{36 R+78} \right. \\& \left.-\sqrt[6]{R}
   \left(-396 (R+2)^{2/3} \sqrt{36 R+78}
   R^{5/6}-90 (R+2)^{2/3} \sqrt{36 R+78}
   R^{11/6}+540 R^3 \right. \right. \\& \left. \left. +3276 R^2  +6591
   R+4394\right)\right)+\frac{\pi ^2 R^{25/6} \sqrt{4
   R+\frac{26}{3}} }{135 \ell_0
   (R+2)^{7/3} (9 R+26)^3}\left(111296640 \sqrt{36 R+78}
   R^{7/6} \right. \\ & \left. +433026 \sqrt{36 R+78} R^{37/6}+5596533
   \sqrt{36 R+78} R^{31/6}+29856168 \sqrt{36 R+78}
   R^{25/6}\right. \\ & \left. +84282948 \sqrt{36 R+78}
   R^{19/6}+132966912 \sqrt{36 R+78}
   R^{13/6}-2598156 (R+2)^{2/3} R^6\right. \\ & \left. -34169688
   (R+2)^{2/3} R^5-186252048 (R+2)^{2/3}
   R^4-539837568 (R+2)^{2/3} R^3\right. \\ & \left. -879420672
   (R+2)^{2/3} R^2-765123840 (R+2)^{2/3}
   R+38667200 \sqrt{36 R+78} \sqrt[6]{R}\right. \\ & \left. 278403840
   (R+2)^{2/3}\right) \epsilon
   +\CO(\epsilon ^2)
   ,
\end{split}
\end{equation*}

\begin{equation*}
\begin{split}
  L =& -\frac{4 }{9 \epsilon
   ^2}\left(\pi ^2 \ell_0^2 R^2 \sqrt{6
   R+13} \left(-6 R+\sqrt[3]{R+2} \sqrt{36 R+78}
   \sqrt[6]{R}-18\right)\right) \\ & +\frac{2 \pi ^2 R^4 \sqrt{6 R+13}}{135
   (R+2)^{5/3} (9 R+26)}
   \left(-3216 \sqrt{36 R+78} R^{7/6}-243 \sqrt{36
   R+78} R^{19/6} \right. \\ & \left.-1524 \sqrt{36 R+78}
   R^{13/6}+1458 (R+2)^{2/3} R^3+9630 (R+2)^{2/3}
   R^2\right. \\ & \left. +21888 (R+2)^{2/3} R-2288 \sqrt{36 R+78}
   \sqrt[6]{R}+17160 (R+2)^{2/3}\right)
   +\CO(\epsilon ^1)
   ,
\end{split}
\end{equation*}
These expressions for $E$, $S$ and $L$ are used for the plot in fig.\ref{fig:ring5dLvsS}.

For the torus in $4+1$ dimensions, the energy, entropy and angular momenta
can be expressed in terms of the derivatives of $T\ln\gpf$ as in \eqref{gpfrot:eq} as well.
Using \eqref{thermexp5dtorus:eq} we find
\begin{equation*}
\begin{split}
  E &=
   \frac{\pi ^3 \ell_0^2 R  \sin (2 \chi_0)}{3\epsilon ^2}
   \left(-R^{1/6} (1 + R)^{1/3} \sqrt{12 + 9 R} (20 + 3 R (8 + 3 R)) \right. \\& \left.
   + 9 (12 + R (18 + R (11 + 3 R)))\right)
   +\frac{\pi ^3 R^3  \csc (2 \chi_0)}{72 (R+1)^{5/3}}
   \left(\sqrt[6]{R} (3 R+4) \sqrt{9 R+12} \left(R \right. \right.
   \\& \left.  \left. (9 R (R (135 R+554)+867)+5564)+1560 \right)
   \right. \\& \left. -3 (R+1)^{2/3} (R (3 R (3 R (3 R (135
   R+734)+4907)+16948)+30658)+7904)\right)
   \\ &
   +\CO\left(\epsilon ^1\right)
   ,
\end{split}
\end{equation*}

\begin{equation*}
\begin{split}
  S &=
   \frac{\pi ^3 \ell_0^2 R^{7/6} \sin (2 \chi_0)}
   {3 \sqrt[3]{R+1} \epsilon^2}
   \left(3 (R+1)^{2/3} \sqrt{9 R+12} (R (3 R+8)+8)\right. \\ & \left.
   -\sqrt[6]{R} (3 R+4) (3 R (3 R+8)+14)\right)
   -\frac{\pi ^3 R^{19/6} (6 R+8)  \csc (2 \chi_0)}{144 \left(\sqrt{3} (R+1)^{7/3}\right)}
   \left(9 (R+1)^{2/3} \right. \\ & \left. \sqrt{3 R+4} (R
   (3 R (R (135 R+554)+891)+2020)+624)\right. \\ & \left.
   -\sqrt{3} \sqrt[6]{R} (3 R+4) (R (9 R (R (135
   R+554)+861)+5440)+1482)\right)
   +\CO\left(\epsilon ^1\right)
   ,
\end{split}
\end{equation*}

\begin{equation*}
\begin{split}
  L_1 &=
   \frac{4 \sqrt{2} \pi ^3 \ell_0^3 R  \sin(\chi_0) \cos^2(\chi_0)}
   {3\epsilon^3}
   \left(3 (R+2) \sqrt{3 R+4}-\sqrt{3} \sqrt[6]{R} \sqrt[3]{R+1} (3 R+4)\right)
   \\ &
   +\frac{\pi ^3 \ell_0 R^3 \sqrt{3 R+4}  \cos (2 \chi_0) \csc
   (\chi_0)}{36 \sqrt{2} (R+1)^{5/3} \epsilon }
   \left(\sqrt[6]{R} (3 R+4) \sqrt{9 R+12} (R (27 R+62)+39) \right.\\ & \left.
   -3 (R+1)^{2/3} (R (3 R (27 R+98)+401)+208)\right)
   +\CO\left(\epsilon ^0\right)
   ,
\end{split}
\end{equation*}

\begin{equation*}
\begin{split}
  L_2 &=
   \frac{4 \sqrt{2} \pi ^3 \ell_0^3 R  \sin^2(\chi_0) \cos(\chi_0)}
   {3\epsilon^3}
   \left(3 (R+2) \sqrt{3 R+4}-\sqrt{3} \sqrt[6]{R} \sqrt[3]{R+1} (3 R+4)\right)
   \\ &
   +\frac{\pi ^3 \ell_0 R^3 \sqrt{3 R+4}  \cos(2\chi_0) \sec(\chi_0)}
   {36 \sqrt{2} (R+1)^{5/3} \epsilon }
   \left(3 (R+1)^{2/3} (R (3 R (27 R+98)+401)+208)\right. \\ & \left.
   -\sqrt[6]{R} (3 R+4) \sqrt{9 R+12} (R (27 R+62)+39)\right)
   +\CO\left(\epsilon ^0\right)
   .
\end{split}
\end{equation*}
We use these expressions for $E$, $S$, $L_1$ and $L_2$ to obtain the
plots in fig.\ref{fig:torus5dLvsS}.

\section{Notation}\label{sec:notation}

We work with the metric signature $(-+++\cdots)$. We use the following index conventions: $\mu,\nu,\ldots$ denote space-time indices and $a,b,\dots$ label planes of rotation, angular velocities, etc. Other symbols used are listed below.

\noindent
\begin{tabular}{rl}
  $a^\mu$ & acceleration \eqref{fluidtensors:eq} \\
  $\mathcal{A}$ & surface area of fluid \\
  $B^m$ & $m$-dimensional ball \\
  $\mathrm{C}_i$ & integration constant \\
  $d$ & dimensionality of fluid \\
  $E$ & total energy of configuration \\
  $\tE$ & see \eqref{thermunits:eq} \\
  $f$ & surface at $f=0$ \\
  $g(\theta)$ & surface at $\rho=g(\theta)$ \\
  $g_i(\theta)$ & terms in $\epsilon$ expansion of $g$ \eqref{scale4d:eq},\eqref{scale5dr:eq},\eqref{scale5dt:eq} \\
  $h_{\mu\nu}$ & induced metric of surface \\
\end{tabular}
\begin{tabular}{rl}
  $h(r_a)$ & surface at $r=h(r_a)$ \\
  $J^\mu_S$ & entropy current \\
  $K^\mu$ & Killing vector for velocity \\
  $L$ & total angular momentum \\
  $\tL$ & see \eqref{thermunits:eq} \\
  $\mathcal{L}$ & distance of fluid from origin \\
  $\ell_i$ & terms in $\epsilon$ expansion of $\mathcal{L}$ \eqref{scalegen:eq},\eqref{scale4d:eq},\eqref{scale5dr:eq},\eqref{scale5dt:eq} \\
  $\mfp$ & thermalisation scale \\
  $l_a$ & rotational Killing vector  \\
  $n$ & no.\ of angular momenta \\
  $n^\mu$ & unit normal of surface \\
\end{tabular}

\noindent
\begin{tabular}{rl}
  $\ploc$ & proper fluid pressure \\
  $P^{\mu\nu}$ & projection tensor \eqref{proj:eq} \\
  $P_a$ & projector for $r_a$ \eqref{projcoord:eq} \\
  $Q_X$ & conserved charge for current $X$ \\
  $q^\mu$ & heat flux \eqref{fluidtensors:eq} \\
  $R$ & radius of fluid tube \\
  $\Ra$ & average radius of fluid tube \eqref{Ravg} \\
  $r$ & radial coordinate \eqref{genmet:eq} \\
  $r_a$ & radial coordinate \eqref{genmet:eq} \\
  $S$ & total entropy of configuration \\
  $\tS$ & see \eqref{thermunits:eq} \\
  $S^m$ & $m$-dimensional sphere \\
  $s$ & proper fluid entropy density \\
  $T$ & overall temperature \\
  $\tT$ & see \eqref{thermunits:eq} \\
  $\tloc$ & proper fluid temperature \\
  $\tc$ & transition temperature \\
  $T^{\mu\nu}$ & stress tensor \\
  $t$ & time coordinate \\
  $u^\mu$ & velocity vector \\
  $w_a$ & scaled angular velocity \eqref{scalegen:eq},\eqref{scale4d:eq},\eqref{scale5dr:eq},\eqref{scale5dt:eq}\\
  $\vec{w}$ & vector field normal to deformed surface \\
  $x_a$ & see \eqref{shift:eq} \\
  $\tilde{x}$ & see \eqref{distunits:eq} \\
  $y(x_a)$ & surface at $r=y(x_a)$ \\
  $\gpf$ & grand partition function \eqref{gpf:eq}  \\
  $\tgpf$ & see \eqref{thermunits:eq} \\
\end{tabular}
\begin{tabular}{rl}
  $\eos$ & equation of state parameter \eqref{bbeos:eq} \\
  $\gamma$ & velocity normalisation $(1-v^2)^{-1/2}$ \\
  $\epsilon$ & expansion parameter \eqref{scalegen:eq},\eqref{scale4d:eq},\eqref{scale5dr:eq},\eqref{scale5dt:eq}\\
  $\zeta$ & bulk viscosity \eqref{extraTvisc:eq} \\
  $\zeta^\mu$ & arbitrary Killing vector \\
  $\eta$ & shear viscosity \eqref{extraTvisc:eq} \\
  $\theta$ & angular coordinate \eqref{newcoord4d:eq},\eqref{newcoord5dring:eq},\eqref{newcoord5dtorus:eq}, \\
  & or coordinate on $S^{d-2n-2}$ \eqref{genmet:eq}\\
  $\vartheta$ & expansion \eqref{fluidtensors:eq} \\
  $\Theta^{\mu\nu}$ & extrinsic curvature of surface \eqref{extrsurf:eq} \\
  $\Theta$ & trace of extrinsic curvature $\Theta^\mu_\mu$ \eqref{trextrsurf:eq} \\
  $\kappa$ & thermal conductivity \eqref{fluidtensors:eq} \\
  $\xi$ & surface thickness \eqref{thick:eq} \\
  $\xi'$ & see \eqref{xipr:eq} \\
  $\rho$ & proper fluid density in \S\eqref{sec:general}\\
   & or shifted radial coordinate \eqref{newcoord4d:eq},\eqref{newcoord5dring:eq},\eqref{newcoord5dtorus:eq} \\
  $\rz$ & plasma vacuum energy \\
  $\rc$ & confinement density \\
  $\sigma$ & surface tension \\
  $\sigma_E$ & surface energy density \\
  $\sigma_S$ & surface entropy density \\
  $\sigma^{\mu\nu}$ & shear tensor \eqref{fluidtensors:eq} \\
  $\phi_a$ & angular coordinate  \eqref{genmet:eq} \\
  $\chi$ & angular coordinate of fluid centre \eqref{newcoord5dtorus:eq} \\
  $\chi_i$ & terms in $\epsilon$ expansion of $\chi$ \eqref{scale5dt:eq} \\
  $\omega_a$ & angular velocity \\
  $\tw_a$ & see \eqref{thermunits:eq} \\
\end{tabular}


\bibliographystyle{utcaps_sl}
\bibliography{plasmaring_new-minimal}

\end{document}

%% file: mydefs.tex
\newcommand{\cdt}{\!\cdot\!}

\newcommand{\prn}[1]{\left ( #1 \right )}

\newcommand{\brk}[1]{\left [ #1 \right ]}

\newcommand{\nrm}[1]{\left\lVert #1 \right\rVert}

\newcommand{\pdiff}[3][\rule{0mm}{0mm}]{\frac{\partial^{#1} #2}{\partial {#3}^{#1}}}

\newcommand{\dr}{\mathrm{d}}
\newcommand{\e}{\mathrm{e}}

\DeclareMathOperator{\Tr}{Tr}

\newlength{\lingap}

%% file: newsymb.tex
\newcommand{\CO}{\mathcal{O}}
\newcommand{\CL}{\mathcal{L}}

\newcommand{\CN}{\mathcal{N}}

\newcommand{\R}{\mathbb{R}}

\newcommand{\p}{\partial}

\newcommand{\half}{\frac{1}{2}}

%% file: plasmaring_new.bbl
\providecommand{\href}[2]{#2}\begingroup\raggedright\begin{thebibliography}{10}

\bibitem{Emparan:2007wm}
R.~Emparan, T.~Harmark, V.~Niarchos, N.~A. Obers, and M.~J. Rodriguez, ``{The
  Phase Structure of Higher-Dimensional Black Rings and Black Holes},''
  \href{http://dx.doi.org/10.1088/1126-6708/2007/10/110}{{\em JHEP} {\bf 10}
  (2007)  110},
\href{http://arxiv.org/abs/0708.2181}{{\tt arXiv:0708.2181 [hep-th]}}.

\bibitem{Niarchos:2008jc}
V.~Niarchos, ``{Phases of Higher Dimensional Black Holes},''
  \href{http://dx.doi.org/10.1142/S0217732308028387}{{\em Mod. Phys. Lett.}
  {\bf A23} (2008)  2625--2643},
\href{http://arxiv.org/abs/0808.2776}{{\tt arXiv:0808.2776 [hep-th]}}.

\bibitem{Maldacena:1997re}
J.~M. Maldacena, ``{The large N limit of superconformal field theories and
  supergravity},'' {\em Adv. Theor. Math. Phys.} {\bf 2} (1998)  231--252,
\href{http://arxiv.org/abs/hep-th/9711200}{{\tt arXiv:hep-th/9711200}}.

\bibitem{Gubser:1998bc}
S.~S. Gubser, I.~R. Klebanov, and A.~M. Polyakov, ``{Gauge theory correlators
  from non-critical string theory},''
  \href{http://dx.doi.org/10.1016/S0370-2693(98)00377-3}{{\em Phys. Lett.} {\bf
  B428} (1998)  105--114},
\href{http://arxiv.org/abs/hep-th/9802109}{{\tt arXiv:hep-th/9802109}}.

\bibitem{Witten:1998qj}
E.~Witten, ``{Anti-de Sitter space and holography},'' {\em Adv. Theor. Math.
  Phys.} {\bf 2} (1998)  253--291,
\href{http://arxiv.org/abs/hep-th/9802150}{{\tt arXiv:hep-th/9802150}}.

\bibitem{Witten:1998zw}
E.~Witten, ``{Anti-de Sitter space, thermal phase transition, and confinement
  in gauge theories},'' {\em Adv. Theor. Math. Phys.} {\bf 2} (1998)  505--532,
\href{http://arxiv.org/abs/hep-th/9803131}{{\tt arXiv:hep-th/9803131}}.

\bibitem{Aharony:2005bm}
O.~Aharony, S.~Minwalla, and T.~Wiseman, ``{Plasma-balls in large N gauge
  theories and localized black holes},''
  \href{http://dx.doi.org/10.1088/0264-9381/23/7/001}{{\em Class. Quant. Grav.}
  {\bf 23} (2006)  2171--2210},
\href{http://arxiv.org/abs/hep-th/0507219}{{\tt arXiv:hep-th/0507219}}.

\bibitem{Lahiri:2007ae}
S.~Lahiri and S.~Minwalla, ``{Plasmarings as dual black rings},''
  \href{http://dx.doi.org/10.1088/1126-6708/2008/05/001}{{\em JHEP} {\bf 05}
  (2008)  001},
\href{http://arxiv.org/abs/0705.3404}{{\tt arXiv:0705.3404 [hep-th]}}.

\bibitem{Bhardwaj:2008if}
S.~Bhardwaj and J.~Bhattacharya, ``{Thermodynamics of Plasmaballs and
  Plasmarings in 3+1 Dimensions},''
\href{http://arxiv.org/abs/0806.1897}{{\tt arXiv:0806.1897 [hep-th]}}.

\bibitem{Emparan:2009cs}
R.~Emparan, T.~Harmark, V.~Niarchos, and N.~A. Obers, ``{Blackfolds},''
\href{http://arxiv.org/abs/0902.0427}{{\tt arXiv:0902.0427 [hep-th]}}.

\bibitem{Andersson:2006nr}
N.~Andersson and G.~L. Comer, ``{Relativistic fluid dynamics: Physics for many
  different scales},''
\href{http://arxiv.org/abs/gr-qc/0605010}{{\tt arXiv:gr-qc/0605010}}.

\bibitem{Son:2007vk}
D.~T. Son and A.~O. Starinets, ``{Viscosity, Black Holes, and Quantum Field
  Theory},''
  \href{http://dx.doi.org/10.1146/annurev.nucl.57.090506.123120}{{\em Ann. Rev.
  Nucl. Part. Sci.} {\bf 57} (2007)  95--118},
\href{http://arxiv.org/abs/0704.0240}{{\tt arXiv:0704.0240 [hep-th]}}.

\bibitem{Caldarelli:2008mv}
M.~M. Caldarelli, O.~J.~C. Dias, R.~Emparan, and D.~Klemm, ``{Black Holes as
  Lumps of Fluid},''
\href{http://arxiv.org/abs/0811.2381}{{\tt arXiv:0811.2381 [hep-th]}}.

\bibitem{Emparan:2008eg}
R.~Emparan and H.~S. Reall, ``{Black Holes in Higher Dimensions},'' {\em Living
  Rev. Rel.} {\bf 11} (2008)  6,
\href{http://arxiv.org/abs/0801.3471}{{\tt arXiv:0801.3471 [hep-th]}}.

\bibitem{Emparan:FW}
J.~Camps, R.~Emparan, and N.~Haddad. {Work in progress}.

\bibitem{Bhattacharyya:2008jc}
S.~Bhattacharyya, V.~E. Hubeny, S.~Minwalla, and M.~Rangamani, ``{Nonlinear
  Fluid Dynamics from Gravity},''
  \href{http://dx.doi.org/10.1088/1126-6708/2008/02/045}{{\em JHEP} {\bf 02}
  (2008)  045},
\href{http://arxiv.org/abs/0712.2456}{{\tt arXiv:0712.2456 [hep-th]}}.

\bibitem{Caldarelli:2008pz}
M.~M. Caldarelli, R.~Emparan, and M.~J. Rodriguez, ``{Black Rings in
  (Anti)-deSitter space},''
  \href{http://dx.doi.org/10.1088/1126-6708/2008/11/011}{{\em JHEP} {\bf 11}
  (2008)  011},
\href{http://arxiv.org/abs/0806.1954}{{\tt arXiv:0806.1954 [hep-th]}}.

\bibitem{Gregory:1993vy}
R.~Gregory and R.~Laflamme, ``{Black strings and p-branes are unstable},''
  \href{http://dx.doi.org/10.1103/PhysRevLett.70.2837}{{\em Phys. Rev. Lett.}
  {\bf 70} (1993)  2837--2840},
\href{http://arxiv.org/abs/hep-th/9301052}{{\tt arXiv:hep-th/9301052}}.

\bibitem{Gregory:1994bj}
R.~Gregory and R.~Laflamme, ``{The instability of charged black strings and
  p-branes},'' \href{http://dx.doi.org/10.1016/0550-3213(94)90206-2}{{\em Nucl.
  Phys.} {\bf B428} (1994)  399--434},
\href{http://arxiv.org/abs/hep-th/9404071}{{\tt arXiv:hep-th/9404071}}.

\bibitem{Caldarelli:2008ze}
M.~M. Caldarelli, O.~J.~C. Dias, and D.~Klemm, ``{Dyonic AdS black holes from
  magnetohydrodynamics},''
\href{http://arxiv.org/abs/0812.0801}{{\tt arXiv:0812.0801 [hep-th]}}.

\bibitem{Maeda:2008kj}
K.-i. Maeda and U.~Miyamoto, ``{Black hole-black string phase transitions from
  hydrodynamics},''
\href{http://arxiv.org/abs/0811.2305}{{\tt arXiv:0811.2305 [hep-th]}}.

\bibitem{Cardoso:2009bv}
V.~Cardoso and O.~J.~C. Dias, ``{Bifurcation of Plasma Balls and Black Holes to
  Lobed Configurations},''
\href{http://arxiv.org/abs/0902.3560}{{\tt arXiv:0902.3560 [hep-th]}}.

\bibitem{Elvang:2007rd}
H.~Elvang and P.~Figueras, ``{Black Saturn},'' {\em JHEP} {\bf 05} (2007)  050,
\href{http://arxiv.org/abs/hep-th/0701035}{{\tt arXiv:hep-th/0701035}}.

\bibitem{Gauntlett:2004wh}
J.~P. Gauntlett and J.~B. Gutowski, ``{Concentric black rings},''
  \href{http://dx.doi.org/10.1103/PhysRevD.71.025013}{{\em Phys. Rev.} {\bf
  D71} (2005)  025013},
\href{http://arxiv.org/abs/hep-th/0408010}{{\tt arXiv:hep-th/0408010}}.

\bibitem{Gauntlett:2004qy}
J.~P. Gauntlett and J.~B. Gutowski, ``{General concentric black rings},''
  \href{http://dx.doi.org/10.1103/PhysRevD.71.045002}{{\em Phys. Rev.} {\bf
  D71} (2005)  045002},
\href{http://arxiv.org/abs/hep-th/0408122}{{\tt arXiv:hep-th/0408122}}.

\bibitem{Iguchi:2007is}
H.~Iguchi and T.~Mishima, ``{Black di-ring and infinite nonuniqueness},''
  \href{http://dx.doi.org/10.1103/PhysRevD.75.064018}{{\em Phys. Rev.} {\bf
  D75} (2007)  064018},
\href{http://arxiv.org/abs/hep-th/0701043}{{\tt arXiv:hep-th/0701043}}.

\bibitem{Evslin:2007fv}
J.~Evslin and C.~Krishnan, ``{The Black Di-Ring: An Inverse Scattering
  Construction},''
\href{http://arxiv.org/abs/0706.1231}{{\tt arXiv:0706.1231 [hep-th]}}.

\bibitem{Elvang:2007hg}
H.~Elvang, R.~Emparan, and P.~Figueras, ``{Phases of Five-Dimensional Black
  Holes},'' {\em JHEP} {\bf 05} (2007)  056,
\href{http://arxiv.org/abs/hep-th/0702111}{{\tt arXiv:hep-th/0702111}}.

\bibitem{Evslin:2008py}
J.~Evslin and C.~Krishnan, ``{Metastable Black Saturns},''
  \href{http://dx.doi.org/10.1088/1126-6708/2008/09/003}{{\em JHEP} {\bf 09}
  (2008)  003},
\href{http://arxiv.org/abs/0804.4575}{{\tt arXiv:0804.4575 [hep-th]}}.

\bibitem{Wald-GeneRela:84}
R.~M. Wald, {\em General Relativity}.
\newblock The University of Chicago Press, Chicago 60637, 1984.

\end{thebibliography}\endgroup
